%% file: main.tex
\pdfoutput=1
\documentclass[12pt,a4paper]{article}

\usepackage{ifthen} 
\newboolean{pdflatex}
\setboolean{pdflatex}{true} 

\newboolean{articletitles}
\setboolean{articletitles}{true} 

\newboolean{uprightparticles}
\setboolean{uprightparticles}{false} 

\def\paperauthors{LHCb collaboration} 
\def\paperasciititle{Machine learning techniques for jet physics at LHCb and application to the search for H to bb and H to cc in s=13$ TeV $pp$ collisions} 
\def\papertitle{Machine learning techniques for jet reconstruction at LHCb and application to the search for $H \to b \Bar{b}$ and $H \to c \Bar{c}$ in $\sqrt{s}=13$ TeV $pp$ collisions} 
\def\paperkeywords{{High Energy Physics}, {LHCb}} 
\def\papercopyright{\the\year\ CERN for the benefit of the LHCb collaboration} 
\def\paperlicence{CC BY 4.0 licence}
\def\paperlicenceurl{https://creativecommons.org/licenses/by/4.0/}

\newif\ifEnableSectionTOCLinks
\EnableSectionTOCLinksfalse 

\input{preamble}
\usepackage{longtable} 

\begin{document}

\renewcommand{\thefootnote}{\fnsymbol{footnote}}
\setcounter{footnote}{1}

\input{title-LHCb-PAPER}


\renewcommand{\thefootnote}{\arabic{footnote}}
\setcounter{footnote}{0}


\cleardoublepage


\pagestyle{plain} 
\setcounter{page}{1}
\pagenumbering{arabic}


\input{body}

\input{acknowledgements}

\newpage
\addcontentsline{toc}{section}{References}
\bibliographystyle{LHCb}
\bibliography{main,standard,LHCb-PAPER,LHCb-CONF,LHCb-DP,LHCb-TDR,LHCb-PUB}

\newpage
\input{Authorship_LHCb-PAPER-2025-034}

\end{document}

%% file: preamble.tex
\usepackage[top=1in, bottom=1.25in, left=1in, right=1in]{geometry}

\usepackage{microtype}
\usepackage{lineno}  
\usepackage{xspace} 
\usepackage{caption} 

\usepackage{graphicx}  
\usepackage{color}
\usepackage{colortbl}
\graphicspath{{./figs/}} 

\usepackage{amsmath} 
\usepackage{amssymb}
\usepackage{amsfonts}
\usepackage{upgreek} 

\newcommand*\patchAmsMathEnvironmentForLineno[1]{%
\expandafter\let\csname old#1\expandafter\endcsname\csname #1\endcsname
\expandafter\let\csname oldend#1\expandafter\endcsname\csname
end#1\endcsname
 \renewenvironment{#1}%
   {\linenomath\csname old#1\endcsname}%
   {\csname oldend#1\endcsname\endlinenomath}%
}
\newcommand*\patchBothAmsMathEnvironmentsForLineno[1]{%
  \patchAmsMathEnvironmentForLineno{#1}%
  \patchAmsMathEnvironmentForLineno{#1*}%
}
\AtBeginDocument{%
\patchBothAmsMathEnvironmentsForLineno{equation}%
\patchBothAmsMathEnvironmentsForLineno{align}%
\patchBothAmsMathEnvironmentsForLineno{flalign}%
\patchBothAmsMathEnvironmentsForLineno{alignat}%
\patchBothAmsMathEnvironmentsForLineno{gather}%
\patchBothAmsMathEnvironmentsForLineno{multline}%
\patchBothAmsMathEnvironmentsForLineno{eqnarray}%
}

\usepackage[pdftex,
            pdfauthor={\paperauthors},
            pdftitle={\paperasciititle},
            pdfkeywords={\paperkeywords}]{hyperref}
\usepackage{hyperxmp}
\hypersetup{
    pdfcopyright={Copyright (C) \papercopyright},
    pdflicenseurl={\paperlicenceurl}
}

\usepackage[colorinlistoftodos,textsize=scriptsize]{todonotes}

\usepackage[bottom,flushmargin,hang,multiple]{footmisc}

\usepackage[all]{hypcap} 

\input{lhcb-symbols-def} 

\hypersetup{
  colorlinks   = true, 
  urlcolor     = blue, 
  linkcolor    = blue, 
  citecolor    = red   
}

\ifEnableSectionTOCLinks
    \usepackage[explicit]{titlesec} 
    
    \let\oldcontentsline\contentsline
    \renewcommand

    \titleformat{\section}{\normalfont\Large\bf}{\hyperlink{tocsection.\thesection}{{\thesection} \parbox[t]{\dimexpr\textwidth-1pc}{#1}}}{1pc}{}

    \titleformat{\subsection}{\normalfont\bf}{\hyperlink{tocsubsection.\thesubsection}{{\thesubsection} \parbox[t]{\dimexpr\textwidth-1pc}{#1}}}{1pc}{}

    \titleformat{name=\section,numberless}[display]{}{}{0pt}{\normalfont\Huge\bfseries #1}
\fi

\usepackage{cite} 
\usepackage{mciteplus}

%% file: lhcb-symbols-def.tex
\usepackage{xspace}
\usepackage{upgreek}

\def\lhcb   {\mbox{LHCb}\xspace}

\def\MagUp {\mbox{\em Mag\kern -0.05em Up}\xspace}

\ifthenelse{\boolean{uprightparticles}}%
{

 \def\PDelta      {\ensuremath{\Delta}\xspace}
 \def\PXi         {\ensuremath{\Xi}\xspace}
 \def\PLambda     {\ensuremath{\Lambda}\xspace}
 \def\PSigma      {\ensuremath{\Sigma}\xspace}
 \def\POmega      {\ensuremath{\Omega}\xspace}
 \def\PUpsilon    {\ensuremath{\Upsilon}\xspace}
 \let\oldPi\Pi
 \def\PPi         {\ensuremath{\oldPi}\xspace}

 \def\PB      {\ensuremath{\mathrm{B}}\xspace}
 \def\PD      {\ensuremath{\mathrm{D}}\xspace}

 \def\PK      {\ensuremath{\mathrm{K}}\xspace}
 \def\Pb      {\ensuremath{\mathrm{b}}\xspace}
 \def\Pc      {\ensuremath{\mathrm{c}}\xspace}

 \def\Ps      {\ensuremath{\mathrm{s}}\xspace}
 \def\Pt      {\ensuremath{\mathrm{t}}\xspace}

 \def\thebaroffset{0.0em}
}
{

 \mathchardef\PDelta="7101
 \mathchardef\PXi="7104
 \mathchardef\PLambda="7103
 \mathchardef\PSigma="7106
 \mathchardef\POmega="710A
 \mathchardef\PUpsilon="7107
 \mathchardef\PPi="7105
 \def\PB      {\ensuremath{B}\xspace}
 \def\PD      {\ensuremath{D}\xspace}

 \def\PK      {\ensuremath{K}\xspace}
 \def\Pb      {\ensuremath{b}\xspace}
 \def\Pc      {\ensuremath{c}\xspace}

 \def\Ps      {\ensuremath{s}\xspace}
 \def\Pt      {\ensuremath{t}\xspace}

 \def\thebaroffset{0.18em}
}
\newcommand{\offsetoverline}[2][\thebaroffset]{\kern #1\overline{\kern -#1 #2}}%

\makeatletter
\ifcase \@ptsize \relax
  \newcommand{\miniscule}{\@setfontsize\miniscule{4}{5}}
\or
  \newcommand{\miniscule}{\@setfontsize\miniscule{5}{6}}
\or
  \newcommand{\miniscule}{\@setfontsize\miniscule{5}{6}}
\fi
\makeatother

\DeclareRobustCommand{\optbar}[1]{\shortstack{{\miniscule (\rule[.5ex]{1.25em}{.18mm})}
  \\ [-.7ex] $#1$}}

\def\squark    {{\ensuremath{\Ps}}\xspace}

\def\cquark    {{\ensuremath{\Pc}}\xspace}
\def\cquarkbar {{\ensuremath{\overline \cquark}}\xspace}
\def\ccbar     {{\ensuremath{\cquark\cquarkbar}}\xspace}
\def\bquark    {{\ensuremath{\Pb}}\xspace}
\def\bquarkbar {{\ensuremath{\overline \bquark}}\xspace}
\def\bbbar     {{\ensuremath{\bquark\bquarkbar}}\xspace}
\def\tquark    {{\ensuremath{\Pt}}\xspace}
\def\tquarkbar {{\ensuremath{\overline \tquark}}\xspace}
\def\ttbar     {{\ensuremath{\tquark\tquarkbar}}\xspace}

\def\KorKbar {\kern \thebaroffset\optbar{\kern -\thebaroffset \PK}{}\xspace}

\def\D       {{\ensuremath{\PD}}\xspace}

\def\DorDbar {\kern \thebaroffset\optbar{\kern -\thebaroffset \PD}\xspace}

\def\Dp      {{\ensuremath{\D^+}}\xspace}
\def\Dm      {{\ensuremath{\D^-}}\xspace}

\def\DpDm    {\ensuremath{\Dp {\kern -0.16em \Dm}}\xspace}

\def\B       {{\ensuremath{\PB}}\xspace}

\def\BorBbar {\kern \thebaroffset\optbar{\kern -\thebaroffset \PB}\xspace}

\def\Bd      {{\ensuremath{\B^0}}\xspace}

\def\BdorBdbar {\kern \thebaroffset\optbar{\kern -\thebaroffset \Bd}\xspace}

\def\Bs      {{\ensuremath{\B^0_\squark}}\xspace}

\def\BsorBsbar {\kern \thebaroffset\optbar{\kern -\thebaroffset \Bs}\xspace}

\def\Y#1S{\ensuremath{\PUpsilon{(#1S)}}\xspace}

\def\LorLbar     {\kern \thebaroffset\optbar{\kern -\thebaroffset \PLambda}\xspace}

\newcommand{\decay}[2]{\mbox{\ensuremath{#1\!\to #2}}\xspace}

\def\to                 {\ensuremath{\rightarrow}\xspace}

\def\AT#1     {\ensuremath{A_{\mathrm{T}}^{#1}}\xspace}           

\def\C#1      {\ensuremath{\mathcal{C}_{#1}}\xspace}                       
\def\Cp#1     {\ensuremath{\mathcal{C}_{#1}^{'}}\xspace}                    
\def\Ceff#1   {\ensuremath{\mathcal{C}_{#1}^{\mathrm{(eff)}}}\xspace}        
\def\Cpeff#1  {\ensuremath{\mathcal{C}_{#1}^{'\mathrm{(eff)}}}\xspace}       
\def\Ope#1    {\ensuremath{\mathcal{O}_{#1}}\xspace}                       
\def\Opep#1   {\ensuremath{\mathcal{O}_{#1}^{'}}\xspace}                    

\newcommand{\nospaceunit}[1]{\ensuremath{\text{#1}}}
\newcommand{\aunit}[1]{\ensuremath{\text{\,#1}}}

\newcommand{\tev}{\aunit{Te\kern -0.1em V}\xspace}
\newcommand{\gev}{\aunit{Ge\kern -0.1em V}\xspace}
\newcommand{\mev}{\aunit{Me\kern -0.1em V}\xspace}
\newcommand{\kev}{\aunit{ke\kern -0.1em V}\xspace}
\newcommand{\ev}{\aunit{e\kern -0.1em V}\xspace}

\newcommand{\mevc}{\ensuremath{\aunit{Me\kern -0.1em V\!/}c}\xspace}
\newcommand{\gevc}{\ensuremath{\aunit{Ge\kern -0.1em V\!/}c}\xspace}
\newcommand{\mevcc}{\ensuremath{\aunit{Me\kern -0.1em V\!/}c^2}\xspace}
\newcommand{\gevcc}{\ensuremath{\aunit{Ge\kern -0.1em V\!/}c^2}\xspace}

\def\mum  {\ensuremath{\,\upmu\nospaceunit{m}}\xspace}

\def\fb   {\ensuremath{\aunit{fb}}\xspace}
\def\invfb   {\ensuremath{\fb^{-1}}\xspace}

\def\gsim{{~\raise.15em\hbox{$>$}\kern-.85em
          \lower.35em\hbox{$\sim$}~}\xspace}
\def\lsim{{~\raise.15em\hbox{$<$}\kern-.85em
          \lower.35em\hbox{$\sim$}~}\xspace}

\def\pt         {\ensuremath{p_{\mathrm{T}}}\xspace}

\def\ptot       {\ensuremath{p}\xspace}

\def\evtgen     {\mbox{\textsc{EvtGen}}\xspace}

\def\geant      {\mbox{\textsc{Geant4}}\xspace}

\def\photos     {\mbox{\textsc{Photos}}\xspace}

\def\pythia     {\mbox{\textsc{Pythia}}\xspace}
\def\fastjet    {\mbox{\textsc{Fastjet}}\xspace}

\def\tell1  {TELL1\xspace}
\def\ukl1   {UKL1\xspace}

\newcommand{\ie}{\mbox{\itshape i.e.}\xspace}

\newcommand{\lhcborcid}[1]{\href{https://orcid.org/#1}{\hspace*{0.1em}\raisebox{-0.45ex}{\includegraphics[width=1em]{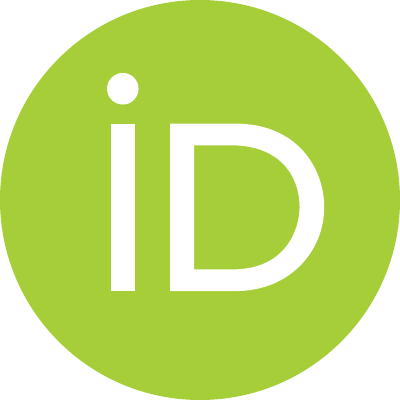}}}}


%% file: title-LHCb-PAPER.tex

\begin{titlepage}
\pagenumbering{roman}

\vspace*{-1.5cm}
\centerline{\large EUROPEAN ORGANIZATION FOR NUCLEAR RESEARCH (CERN)}
\vspace*{1.5cm}
\noindent
\begin{tabular*}{\linewidth}{lc@{\extracolsep{\fill}}r@{\extracolsep{0pt}}}
\ifthenelse{\boolean{pdflatex}}
{\vspace*{-1.5cm}\mbox{\!\!\!\includegraphics[width=.14\textwidth]{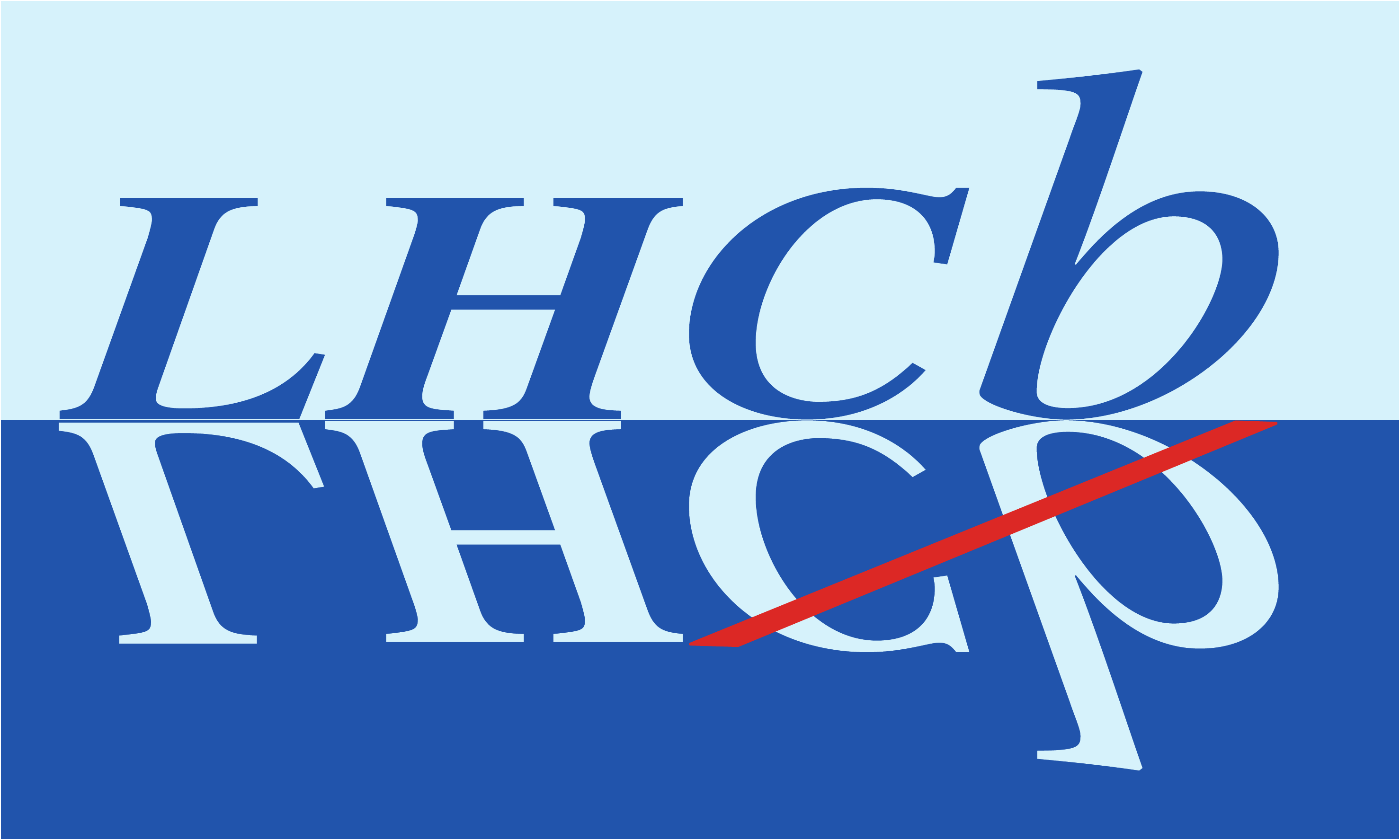}} & &}%
{\vspace*{-1.2cm}\mbox{\!\!\!\includegraphics[width=.12\textwidth]{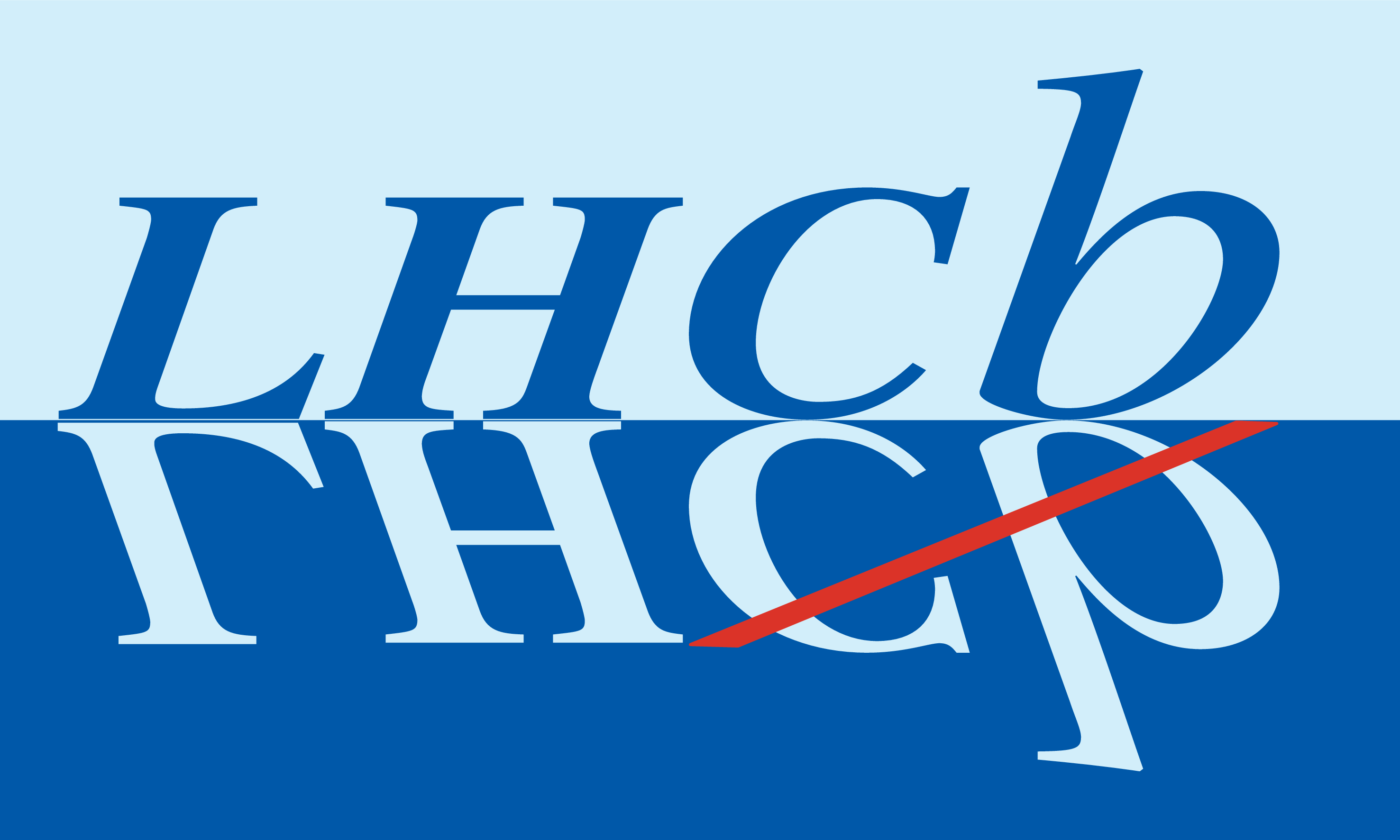}} & &}%
\\
 & & CERN-EP-2025-275 \\  
 & & LHCb-PAPER-2025-034 \\  
 & & \today \\ 
 & & \\
\end{tabular*}

\vspace*{1.0cm}

{\normalfont\bfseries\boldmath\huge
\begin{center}
  \papertitle 
\end{center}
}

\vspace*{1.0cm}

\begin{center}
\paperauthors\footnote{Authors are listed at the end of this paper.}
\end{center}

\vspace{\fill}


\begin{abstract}
  \noindent
  Two machine learning techniques for jet measurements at the LHCb experiment are presented: a regression-based method for jet-energy calibration and a deep neural network algorithm for jet flavour tagging, distinguishing between $\bquark$-quark, $\cquark$-quark, and light parton jets. These techniques are applied to a search for inclusive $H \to \bbbar$ and $H \to \ccbar$ decays using a LHCb dataset corresponding to an integrated luminosity of 1.6$\invfb$. The observed (expected) $95\%$ confidence level upper limits correspond to 6.6 (11.1) times the SM cross-section for the $H \to \bbbar$ process, and 1003 (1834) times the SM cross-section for the $H \to \ccbar$ process.

\end{abstract}

\vspace*{1.0cm}

\begin{center}
  Published in Journal of High Energy Physics 07 (2026) 276
\end{center}

\vspace{\fill}

{\footnotesize 
\centerline{\copyright~\papercopyright. \href{\paperlicenceurl}{\paperlicence}.}}
\vspace*{2mm}

\end{titlepage}


\newpage
\setcounter{page}{2}
\mbox{~}
%
%
%
%

%% file: body.tex
\section{Introduction}
\label{sec:Introduction}
Machine learning has found widespread application in high-energy physics~\cite{ml_hep}, with jet physics representing one of its most successful domains~\cite{ml_jets}. The ATLAS and CMS collaborations have demonstrated that the performance of jet reconstruction and identification can be substantially enhanced by using machine learning techniques~\cite{ml_jets_atlas, ml_jets_cms}. These techniques exploit the jet substructure, \emph{i.e.}, the particles forming the jets, measured as collections of tracks and calorimeter clusters. Secondary vertices and hadrons reconstructed from their decay products can also be included in the jet substructure~\cite{jet_substructure}. 
Features related to these objects are used as inputs to machine learning models, which typically fall within the domain of deep learning. 
Successful models applied to jet physics include deep neural networks~\cite{deepjet}, graph neural networks~\cite{gnn_atlas}, and ParticleNet~\cite{particlenet}, although many others have been tested with varying performance. 
The machine learning tasks considered range from jet-energy reconstruction to jet-flavour identification. 
An important application of machine learning techniques concerns the so-called fat jets, \emph{i.e.}, jets with a large radius, where two or more jets may merge into a single jet~\cite{fatjet}. In these cases, their classification---through the identification of the merged objects and/or the decay process that produced them---can be effectively addressed using deep learning methods. 

One of the most notable applications has been the search for the Higgs boson decay into charm quarks.
Thanks to machine learning and advanced analysis techniques, the sensitivity of this search has been extended beyond the reach of conventional methods~\cite{VH_atlas, VH_cms, HccCMS, ttH_cc_cms}, thereby opening the prospect of its observation with the upgraded detectors in the High-Luminosity LHC era.
In particular, the ATLAS and CMS experiments have searched for the $H \to \ccbar$ decay in the $VH$ production mode, where $V$ is a $W$ or $Z$ boson, setting observed upper limits of 12 and 14 times the Standard Model (SM) prediction at the 95\% confidence level (CL)~\cite{VH_atlas, VH_cms}.  
The CMS collaboration has also searched in the inclusive topology, targeting boosted Higgs bosons reconstructed as a single large-radius jet, and obtained a limit of 47 times the SM prediction~\cite{HccCMS}.  
More recently, the CMS collaboration reported a preliminary result in the $\ttbar H$ channel, with an observed upper limit of 8 times the SM expectation~\cite{ttH_cc_cms}. Further advances in jet physics are expected through the use of cutting-edge machine learning technologies.

The LHCb experiment has also employed machine learning techniques in jet physics. The first application utilised a boosted decision tree classifier~\cite{Breiman,AdaBoost} for jet-flavour classification, exploiting variables related to reconstructed secondary vertices~\cite{LHCb-PAPER-2015-016}. A related development was a variational quantum machine learning circuit for the classification of \bquark- and \bquarkbar-jets~\cite{QML_LHCb}. Additional machine learning methods have been applied to separate signals with jets in the final state from background processes~\cite{LHCb-PAPER-2016-038, LHCb-CONF-2016-006}.
These techniques have already been employed to measure differential cross-sections for \bbbar and \ccbar dijet production~\cite{dijet}. Furthermore, the LHCb collaboration has demonstrated its ability to extract the $Z \to \bbbar$ signal from the dominant QCD multijet background~\cite{LHCb-PAPER-2017-024}. Using part of the Run~1 data, corresponding to the dataset collected in 2012 at $\sqrt{s}=8$ TeV, the LHCb collaboration conducted for the first time a search for the $\decay{VH}{\cquark \cquarkbar}$ process, setting an upper limit of 6400 times the SM expectation~\cite{LHCb-CONF-2016-006}.
In this paper, two novel methods are presented, aimed at improving LHCb jet reconstruction and identification performance:
\begin{enumerate}
    \item a regression-based jet-energy correction designed to improve the Higgs dijet invariant-mass resolution; and
    \item a deep neural network (DNN)-based jet identification algorithm that exploits full jet substructure information to discriminate between $\bquark$, $\cquark$, and light jets.\footnote{Throughout this work, ``light jets'' refers to jets initiated by $u$, $d$, or $s$ quarks, as well as gluons.}
\end{enumerate}
The improvements in the dijet invariant-mass resolution and in jet identification efficiencies achieved with these techniques, with respect to the standard LHCb algorithms, are also evaluated.

The effectiveness of the machine learning techniques presented in this paper is demonstrated in a search for inclusive $H \to \bbbar$ and $H \to \ccbar$ decays in the dijet final state, using a dataset of pp collisions collected in 2016 corresponding to an integrated luminosity of 1.6\invfb.
Unlike previous LHCb analyses, which focused on vector boson associated production, the study in this paper targets inclusive $H \to \bbbar$ and $H \rightarrow \ccbar$ processes.
In the inclusive search, no assumptions are made on the Higgs production mechanism, making it suitable for comparisons with several different BSM models~\cite{hadroproduction}. 
A major challenge in the inclusive search is the modelling of the multijet QCD background. A data-driven approach, based on the output of the DNN score (usually referred to as ``DNN tagging"), is used to describe this background.
Upper limits are set on the $H \to \bbbar$ and $H \to \ccbar$ production cross-sections, and the limit on the $H \to \ccbar$ cross-section is used to derive a limit on the charm Yukawa coupling ($y_c$).

The paper is structured as follows: Sec.~\ref{sec:Detector} provides an overview of the LHCb detector, Sec.~\ref{sec:JetReconstructionAndEventReconstruction} details the techniques used for particle and jet reconstruction, including the particle flow algorithm and jet clustering methods. In Sec.~\ref{sec:regression}, a regression-based approach for jet-energy correction is introduced, aiming to improve the dijet invariant-mass resolution. Section~\ref{sec:JetFlavourIdentification} discusses the methods employed for jet flavour tagging, including secondary-vertex reconstruction and a DNN-based classification algorithm. The search strategy for the $H \rightarrow \bbbar$ and $H \rightarrow \ccbar$ decays is presented in Sec.~\ref{sec:SearchStrategy}, covering event selection, background estimation, and statistical analysis techniques. Section~\ref{sec:SystematicUncertainties} describes the systematic uncertainties affecting the measurement, while Sec.~\ref{sec:ResultsAndProspects} presents the results obtained from the analysis and their implications. Finally, Sec.~\ref{sec:Summary} summarises the findings and outlines prospects for future studies.

\section{Detector and simulation}
\label{sec:Detector}

The \lhcb detector~\cite{Alves:2008zz,LHCb-DP-2014-002} is a single-arm forward spectrometer covering the \mbox{pseudorapidity} range $2<\eta <5$, designed for the study of particles containing \bquark or \cquark quarks. The detector used to collect the data analysed in this paper includes a high-precision tracking system consisting of a silicon-strip vertex detector surrounding the $pp$ interaction region~\cite{LHCb-DP-2014-001}, a large-area silicon-strip detector located
upstream of a dipole magnet with a bending power of about $4{\mathrm{\,T\,m}}$, and three stations of silicon-strip detectors and straw drift tubes~\cite{LHCb-DP-2017-001} placed downstream of the magnet.
The tracking system provides a measurement of the momentum, \ptot, of charged particles with a relative uncertainty that varies from 0.5\% at low momentum to 1.0\% at 200\gevc. The minimum distance of a track to a primary vertex (PV), the impact parameter (IP), is measured with a resolution of $(15+29/\pt)\mum$,
where \pt is the component of the momentum transverse to the beam, in\,\gevc.
Different types of charged hadrons are distinguished using information from two ring-imaging Cherenkov detectors~\cite{LHCb-DP-2012-003}. Photons, electrons, and hadrons are identified by a calorimeter system consisting of scintillating-pad and preshower detectors, an electromagnetic and a hadronic calorimeter. Muons are identified by a system composed of alternating layers of iron and multiwire proportional chambers~\cite{LHCb-DP-2012-002}. The online event selection is performed by a trigger~\cite{LHCb-DP-2012-004}, which consists of a hardware stage, based on information from the calorimeter and muon systems, followed by a software stage, which applies a full event reconstruction.
At the hardware trigger stage, events for this analysis are required to contain one or two reconstructed muons with high \pt or a reconstructed hadron, photon, or electron with high transverse energy in the calorimeters. A global event cut (GEC) on the number of hits in the scintillating-pad detector (nSPD) is also applied. The software trigger has a first stage that requires at least one or two tracks compatible with the decay of a $\bquark$ or $\cquark$ hadron. The second stage of the software trigger requires two reconstructed jets with $\pt > 17 \gevc$ and a secondary vertex in the jet cone. Details on the jet reconstruction are given in the next section. Triggered data further undergo a centralised, offline processing step to deliver physics-analysis-ready data across the entire \lhcb physics programme~\cite{Stripping}.

Simulation is required to model and correct for the effects of the detector acceptance and resolution, and the imposed selection requirements. In the simulation, $pp$ collisions are generated using \pythia~\cite{Sjostrand:2007gs,*Sjostrand:2006za} with a specific \lhcb configuration~\cite{LHCb-PROC-2010-056}.  Decays of unstable particles are described by \evtgen~\cite{Lange:2001uf}, in which final-state radiation is generated using \photos~\cite{Golonka:2005pn}. The interaction of the generated particles with the detector, and its response, are implemented using the \geant toolkit~\cite{Allison:2006ve, *Agostinelli:2002hh} as described in Ref.~\cite{LHCb-PROC-2011-006}.

\section{Particle and jet reconstruction}
\label{sec:JetReconstructionAndEventReconstruction}
Event reconstruction is based on a particle flow algorithm~\cite{LHCb-PAPER-2013-058}, which combines information from all subdetectors to identify and reconstruct individual particles, namely charged hadrons, photons, neutral hadrons, electrons, and muons. Each reconstructed particle, referred to as a particle flow object, is obtained by associating tracks and calorimeter clusters, using particle identification information to ensure consistent topologies. To take advantage of the excellent momentum resolution provided by the tracking system, reconstructed tracks are used as the charged particle inputs to the jet reconstruction. Neutral particle inputs are obtained from energy deposits in the electromagnetic and hadronic calorimeters. When these deposits are matched to tracks, the expected calorimeter response for the corresponding charged particles is subtracted. This expected energy is estimated using particle identification information to determine whether the track is likely to be a charged hadron, muon, or electron. If a significant amount of energy remains after this subtraction, it is attributed to a neutral particle detected in the calorimeter. Primary vertices are reconstructed by clustering tracks that originate from a common point along the beam line, using a vertex-fitting algorithm to determine the position with optimal spatial resolution~\cite{Kucharczyk:1756296}.

Jets are reconstructed using particle flow objects as input. The objects are combined employing the anti-$k_{\rm t}$ algorithm~\cite{antikt}, as implemented in the \fastjet software package~\cite{fastjet}, with a jet radius parameter of $R=0.5$. To improve the rejection of fake jets, such as jets originating from combinatorial background and isolated high-energy leptons, additional criteria, similar to those explained in Ref.~\cite{LHCb-PAPER-2013-058}, are imposed. In particular, jets are required to contain at least two charged particles matched to the same PV, at least one track with $\pt>1.2 \gevc$, no single particle carrying more than $75\%$ of the jet $\pt$, and to have the fraction of the jet $\pt$ carried by charged particles greater than $10\%$. These requirements have been optimised using simulated samples produced with 2016 running conditions.

\section{Regression techniques for jet-energy correction}
\label{sec:regression}

Compared to jets from light-flavour quarks or gluons, heavy-flavour (\bquark or \cquark) jets require dedicated energy corrections due to unique features. These jets often involve semileptonic decays with undetected neutrinos, leading to energy underestimation and degraded resolution.
At the LHCb experiment, the standard correction method applies a multiplicative factor to the reconstructed jet four-momentum defined as

\begin{equation}
k_{\mathrm{truth}}\equiv \frac{E_{\mathrm{truth}}}{E_{\text {reco }}} \ ,
\end{equation}
where $E_{\mathrm{truth}}$ is the truth-level energy of the jet, and $E_{\text {reco }}$ is the reconstructed-level energy of the jet. 
This factor is modelled as a cubic function of $\pt$, $\eta$, $\phi$, charged-particle fraction (CPF), and the number of reconstructed PVs (nPVs). While designed to match the reconstructed and truth-level invariant masses, this jet-energy correction (JEC) method leads to a slight degradation in resolution and a reduction in sensitivity compared with uncorrected jets.

In this section, a new regressor technique to correct jet energy is presented. This new regressor receives characteristics of reconstructed jets as input variables, while the target variable is the truth-level energy of jets. The regression is done separately for the leading and subleading jets, \ie the jet with the highest transverse momentum and the jet with the second-highest transverse momentum in the event, respectively.

The correction function is estimated using a gradient boosting regressor
(GBR) implemented in the \texttt{scikit-learn} library~\cite{Scikit-learn-paper}. After establishing the regressor, the input variables, and the target, the hyperparameters are optimised.

The regression inputs provide information about the kinematics, shape, and composition of the reconstructed jets, which can be correlated with the dijet invariant-mass difference $\Delta m \equiv m_{\mathrm{reco}} - m_{\mathrm{truth}}$, where $m_{\mathrm{reco}}$ is the reconstructed, and $m_{\mathrm{truth}}$ is the true invariant mass.
The following inputs are considered:
\begin{itemize}
\item  Jet kinematics: \pt, $\eta$ and invariant mass prior to any JEC;
\item  Jet composition: number of constituents, the largest \pt and energy values among the constituents, and the charged-particle fraction;
\item Jet-energy distribution in rings of $\Delta R \equiv \sqrt{(\Delta \eta)^2 + (\Delta \phi)^2}$ around the jet axis (\mbox{$\Delta R \in$ 0.00–0.05}, 0.05–0.10, 0.1–0.2, 0.2–0.3, 0.3–0.4), where $\Delta \eta$ and $\Delta \phi$ are the pseudorapidity and azimuthal angle differences between a constituent particle momentum vector and the jet axis, respectively;
\item Information about muons: number of muons in the jet, their energy, and momentum;
\item Maximum and minimum energy deposited in the ECAL by neutral particles in the jet;
\item Number of tracks in the jet and number of reconstructed PVs in the event.
\end{itemize}

To improve the regressor's performance, six hyperparameters are considered: the loss function, the number of estimators, the maximum depth of each estimator, the minimum number of samples per leaf, the minimum number of samples in each node required to do a split, and the learning rate. They are tuned using the halving grid search cross-validation algorithm~\cite{Scikit-learn-paper}. 
Once the optimal hyperparameters have been determined, two models are obtained by applying the regressor to both leading and subleading reconstructed jets in a dataset consisting of simulated $H \rightarrow \bbbar$ decays. 

The first step is to divide the dataset into training and testing samples. Dijet events are arbitrarily assigned to one sample or the other, ensuring that the distributions of relevant features are similar in both. This procedure helps to minimise differences between the training and testing datasets and reduces potential biases during model evaluation. The training set also includes simulated signal events, combined with background events, to provide a representative sample for learning. The datasets used in the training are 35\,000 simulated $H\rightarrow \bbbar$ events as signal and 46\,000 events from the dijet data sample as background. Once the training phase is completed, both models are requested to predict the energy $E^i_{{\rm GBR}}$ of the leading ($i=0$) and the subleading ($i=1$) jets. Next, the correction factor between the initial reconstructed energy $E^i_\text{reco}$ without the prior correction and the energy resulting from the regression $E^i_\text{GBR}$ is computed as
\begin{equation}
k^{i}_\text{GBR}=\frac{E^i_\text{GBR}}{E^i_\text{reco}}, \hspace{0.8cm} i=0,1 .
\end{equation}
\begin{figure}
    \centering
    \includegraphics[scale=0.5]{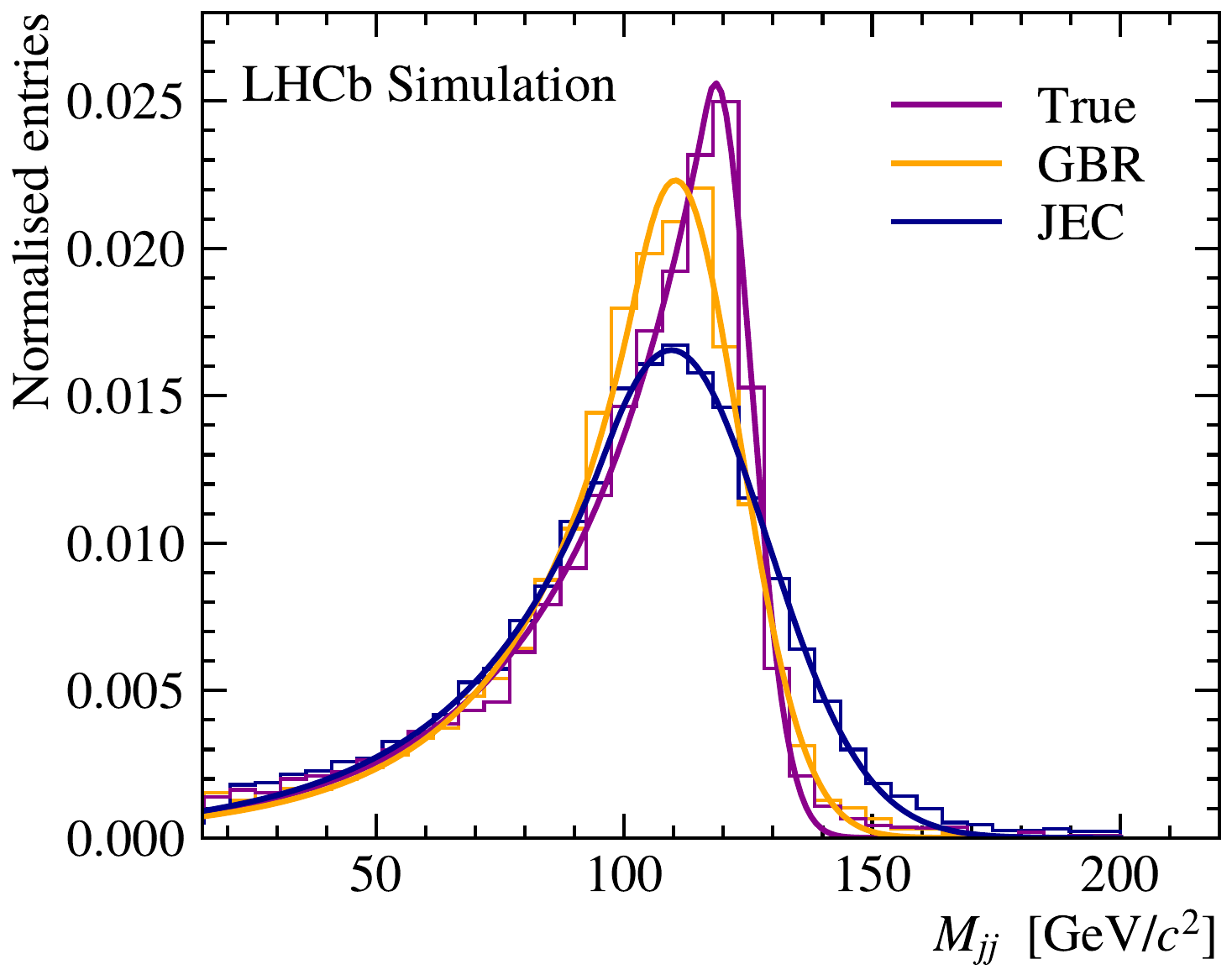}
    \caption{Normalised distributions of the $\bbbar$ invariant mass in a $H \to \bbbar$ simulation sample, obtained with truth-level jets and reconstructed-level jets with default JEC correction and GBR correction. The results of the fit to these distributions with Crystal Ball functions are also shown. }
    \label{fig:regression_fit}
\end{figure}
The correction factors $k^i_\text{GBR}$ are applied to all components of momentum equally, as $\eta$ and $\phi$ are assumed to be the same for reconstructed-level and truth-level jets. Once the four-momenta of both jets are corrected, the invariant mass is recalculated. 
The same factors $k^i_\text{GBR}$ are also used in the data to correct the jet energy. Figure~\ref{fig:regression_fit} shows the comparison between true and reconstructed $\bbbar$ dijet invariant mass in simulated events, using both the standard JEC method and the new GBR algorithm just described. The improvement in the dijet invariant-mass resolution is evident. Several cross-checks have been performed to demonstrate that the GBR technique is unbiased as a function of the dijet mass, enabling its use in other use cases. Particularly, no bias for the jet $\pt$ resolution is found as a function of the dijet invariant mass.

\section{Jet flavour identification}
\label{sec:JetFlavourIdentification}
In this section, the techniques used to identify the flavour of the quarks generating the jets are described. First, the secondary-vertex tagging algorithm~\cite{LHCb-PAPER-2015-016}, used as the standard for heavy-flavour jet measurements at LHCb, is introduced. Then, the DNN algorithm specifically developed for this analysis is presented.

\subsection{Secondary-vertex tagging}
\label{subsec:SecondaryVertexTagging}

Within the context of jet heavy-flavour tagging at LHCb, jets undergo an algorithm which is referred to as ``secondary-vertex tagging''. The secondary-vertex tagging algorithm reconstructs secondary vertices using tracks inside and outside of the jet and is described in detail in Ref.~\cite{LHCb-PAPER-2015-016}. In this algorithm, tracks that have a significant \pt and displacement from every PV are combined to form two-body secondary vertices. Then, good-quality two-body secondary vertices are linked together if they share one track, in order to form $n$-body secondary vertices.
If a secondary vertex is found inside the cone of the jet, the jet is tagged as likely to be originating from $\bquark$- or $\cquark$-quark fragmentation. The secondary-vertex tagging efficiency, determined in simulated samples described in Sec.~\ref{sec:DataSim}, is about 60\% for $\bquark$-jets and 20\% for $\cquark$-jets. These values are lower than those obtained in the $\sqrt{s}=7$ and $8$~\tev datasets studied in Ref.~\cite{LHCb-PAPER-2015-016}. The relative efficiency loss is below the 10\% level, and is explained by the higher particle multiplicity of $\sqrt{s}=13$ \tev events with respect to $\sqrt{s}=7$ and $8$~\tev events, which introduces more combinatorial background in the secondary-vertex finding.

\subsection{Deep neural networks for flavour identification}
\label{subsec:DeepNeuralNetwork}
In previous LHCb analyses, to further distinguish light-flavour jets from heavy-flavour jets and $\bquark$-jets from $\cquark$-jets, multivariate analysis algorithms have been used, as described in Ref.~\cite{LHCb-PAPER-2015-016}. In this paper, a new approach based on deep neural networks (DNNs) is presented. The DNN used in this study resembles the \emph{DeepJet} algorithm~\cite{deepjet} developed by the CMS collaboration. 
Particularly, the jet substructure features are grouped into four categories: \emph{charged particles}, \emph{neutral particles}, \emph{secondary vertex}, and \emph{global}.
For the \emph{charged particles} and \emph{neutral particles} categories, the inputs include kinematic and reconstruction information from tracks and calorimeter clusters not associated with a track, as well as particle identification variables. The \emph{secondary vertex} category includes kinematic properties and other features of a reconstructed secondary vertex within the jet. The \emph{global} category contains jet-level kinematic and shape variables, such as the $\pt$,  $\eta$, invariant mass, and jet radius.
The total number of input features is fixed to 421. A maximum of $20$ charged particles (ordered by decreasing impact parameter) and $10$ neutral particles (ordered by decreasing energy) per jet are considered. If the number of particles in a given jet is insufficient to fill all input slots, the remaining features are set to zero. An important advantage of the DNN architecture is that it can be applied even when no secondary vertex is reconstructed; in such cases, the secondary-vertex features are set to zero.

The DNN structure consists of several sequential steps. First, all input features are normalised for effective network processing. Next, a 1D convolutional layer~\cite{Lee:2019cad} is applied exclusively to the charged particles and neutral particles features. This is followed by a recursive layer, specifically a long short-term memory layer~\cite{PhysRevD.94.112002}, also applied only to the charged and neutral features. Finally, a dense layer is applied to all features, producing the network output.
The output of the DNN algorithm is a set of three probabilities, namely $P_b$, $P_c$, and $P_q$, which represent the probability for a jet to originate from a $\bquark$, $\cquark$, or light quark, respectively.

Simulated samples of $\bbbar$, $\ccbar$, and light dijets are used for training and testing of the DNN. The total dataset is divided into training, evaluation, and testing subsets by randomly selecting jets from the simulated samples, with $50\%$ allocated for training, $20\%$ for evaluation, and $30\%$ for testing. The reconstructed jets are matched with truth-level jets. It is also required that the jet pseudorapidity must be in the range $2.2 < \eta < 4.2$ and the jet transverse momentum must be greater than $10\gevc$. The reduced $\eta$ range with respect to the full LHCb acceptance is chosen to ensure a well-understood jet reconstruction and identification efficiency. 
\begin{figure}[t]
    \centering
    \includegraphics[width=0.49\textwidth]{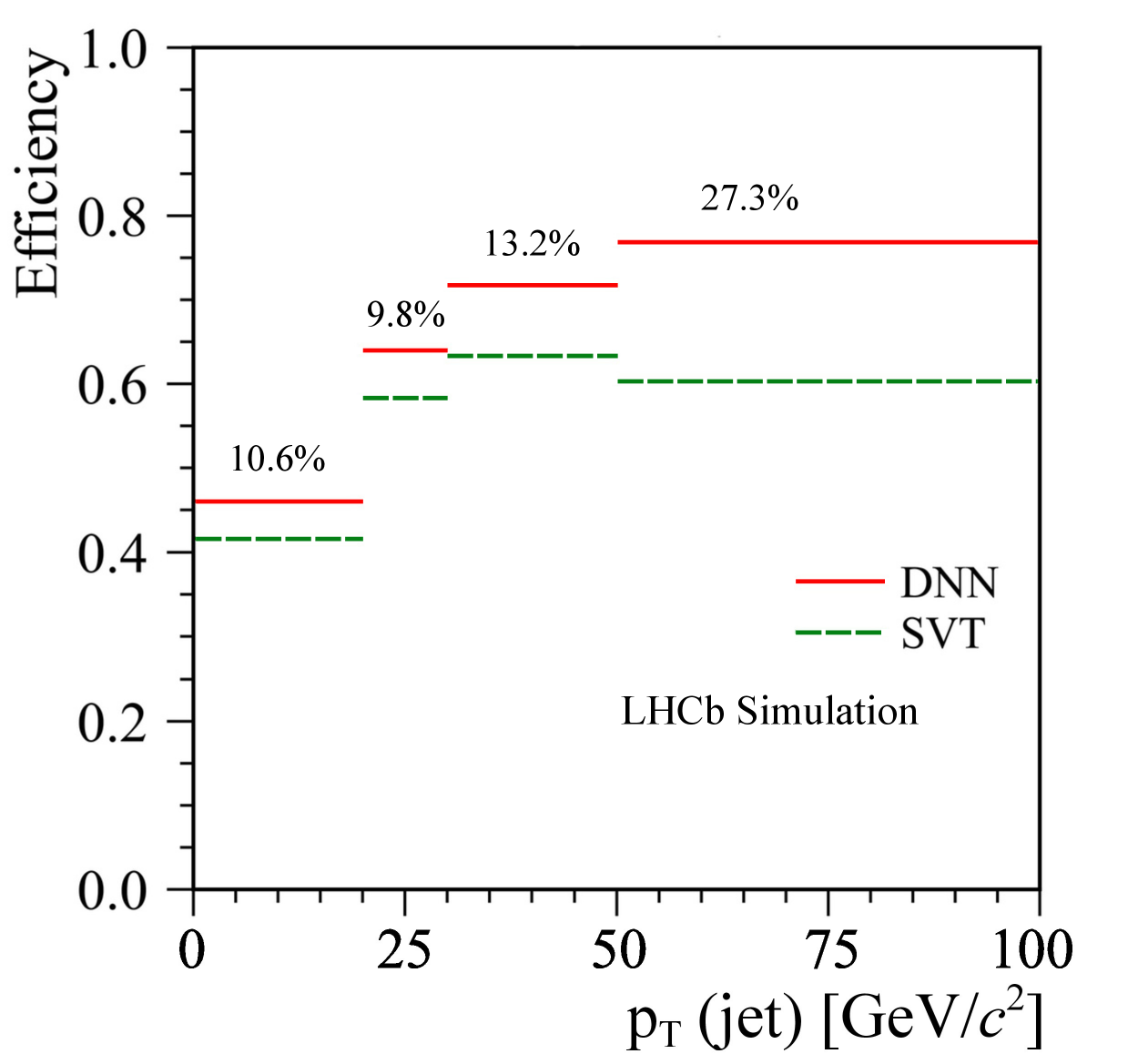}
    \includegraphics[width=0.49\textwidth]{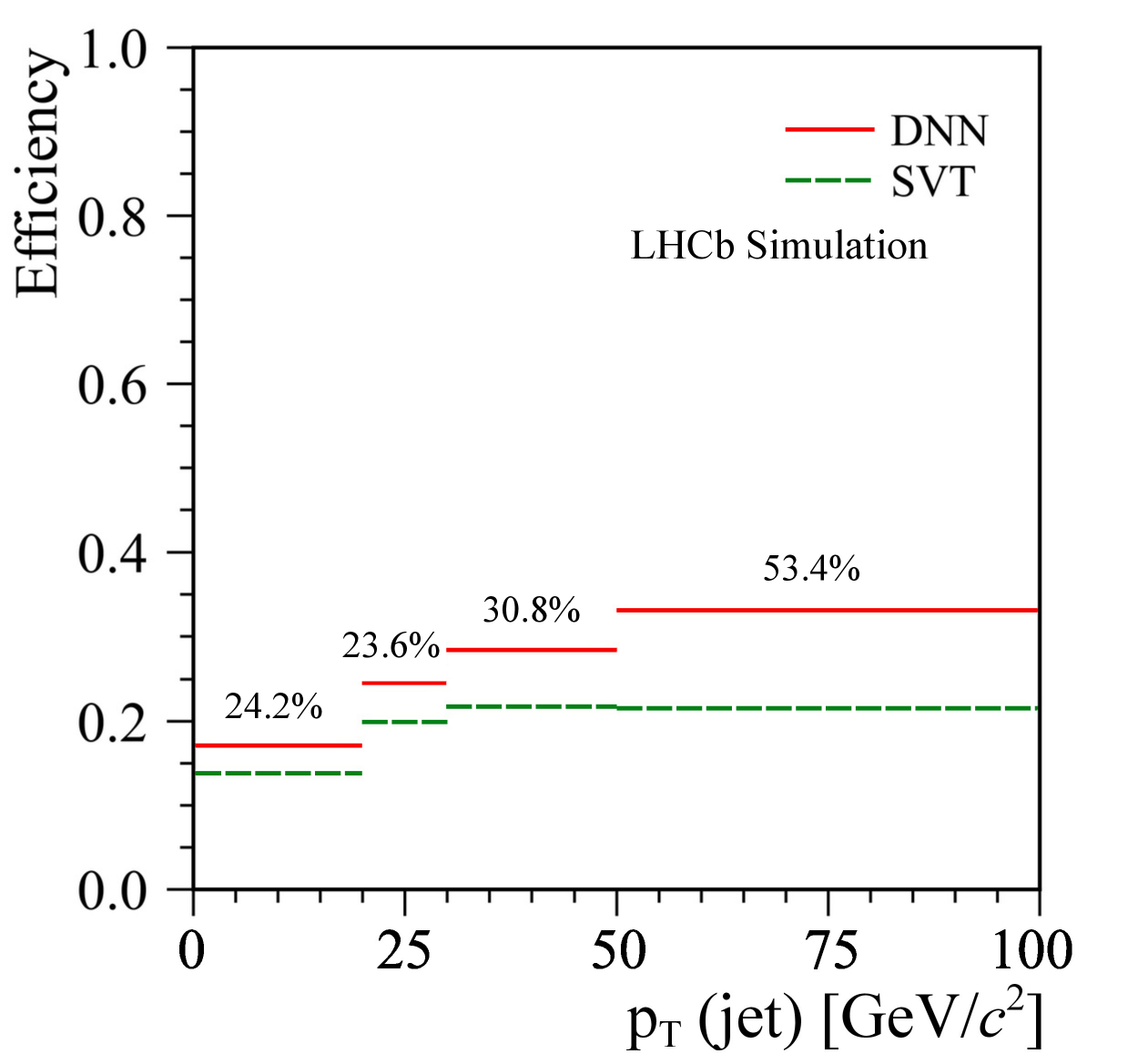}
    \caption{Single-jet heavy-flavour tagging efficiency $\varepsilon_{\mathrm{tag}}$ for DNN and secondary-vertex tagging (SVT), in the (left) $\bbbar$ sample and (right) $\ccbar$ sample, as a function of jet $\pt$. For each bin, the relative improvement of the DNN over the SVT algorithm is indicated.}
    \label{fig:dnn_eff}
\end{figure}
To evaluate the DNN performance, the algorithm is applied to $\bbbar$, $\ccbar$, and light dijet samples that have passed the selection requirements described above. 
The comparison between the DNN and the secondary-vertex tagging algorithm is performed in terms of tagging efficiency $\varepsilon_{\mathrm{tag}}$, defined as
\begin{equation}
    \varepsilon_{\mathrm{tag}}= \frac{N_{\mathrm{tag}}}{N_{\mathrm{rec}}}\ ,
\end{equation}
where $N_{\mathrm{rec}}$ is the number of reconstructed jets, and $N_{\mathrm{tag}}$ the number of those tagged as $\bquark$- or $\cquark$-jets. To have a fair comparison between DNN and secondary-vertex tagging, the requirements on the DNN probabilities are optimised in order to have the same light-jet misidentification rate as the secondary-vertex tagging algorithm (close to $1 \%$). Figure~\ref{fig:dnn_eff} shows the comparison between DNN and secondary-vertex tagging efficiencies for identifying a $\bquark$- or a $\cquark$-jet, as a function of the jet $\pt$. For each bin of jet $\pt$, the relative efficiency improvement coming from the DNN with respect to secondary-vertex tagging is also shown. For $\bquark$-jets, the improvement is greater than $9\%$, while for the $\cquark$-jets, this improvement is always above $20\%$.

\section{Application to \boldmath{$H \to \bbbar$} and \boldmath{$H \to \ccbar$} searches}
\label{sec:SearchStrategy}

In this section, the search for the decays $H \to \bbbar$ and $H \to \ccbar$ at LHCb is presented as the first application of the machine learning techniques described in this paper. The analysis is performed in an inclusive final state, \ie without selecting a specific Higgs production mechanism, and focuses on events containing two resolved jets identified as $\bquark$- or $\cquark$-jets, using the techniques described in Sec.~\ref{sec:JetFlavourIdentification}. The primary observable used in the search is the dijet invariant mass.

The analysis strategy is based on selecting a dijet event sample enriched in $H \to \bbbar$ or $H \to \ccbar$ decays.
Signal regions (SRs) are defined by applying thresholds on the tagging DNN scores, while orthogonal control regions (CRs), composed of mixed-flavour jet pairs (such as $b\ell$, $bc$, and $c\ell$, where $\ell$ is a light jet), are used to constrain the multijet QCD background. The background is modelled using data in the CRs, together with a transfer function ($\mathrm{TF}$) that accounts for kinematic differences between the CRs and SRs.

A comprehensive background model is constructed, incorporating contributions from multijet QCD, $Z \to \bbbar$, $Z \to \ccbar$, top quark decays, and hadronic $W$ decays. A simultaneous fit to the dijet mass spectrum in the SR is performed, with the signal and multijet QCD components left free to vary. The $Z$ boson background sources are included according to SM expectations, but with an overall normalisation factor allowed to vary. A transfer function is included in the model to capture residual differences between control and signal regions, and its parameters are determined in the fit.

The upper limits on the signal yields and production cross-sections for $H \to \bbbar$ and $H \to \ccbar$ decays are determined using the CL$_\mathrm{s}$ method~\cite{CLs}. Separate analysis paths are followed for the two decay modes, each with independent definitions of the SRs and CRs, fit configurations, and systematic uncertainty evaluations.

\subsection{Data and simulation samples}
\label{sec:DataSim}

The data sample used to extract upper limits on the \decay{H}{\bbbar} and \decay{H}{\ccbar} decays is selected by a second-level software trigger line that requires two jets, each with \pt greater than $17 \gevc$, and a reconstructed secondary vertex within the jet cone, as described in Sec.~\ref{sec:Detector}. 

Dedicated simulated samples of the \decay{H}{\bbbar} and \decay{H}{\ccbar} processes have been generated to study the signal characteristics. These samples are used to determine selection efficiencies and to model the signal dijet invariant-mass distribution, and are also used to evaluate the performance of the GBR algorithm described in Sec.~\ref{sec:regression}. Simulated samples of $\bbbar$, $\ccbar$, and light dijets are used to evaluate the performance of the DNN, as explained in Sec.~\ref{sec:JetFlavourIdentification}. Simulated samples of \decay{Z}{\bbbar}, \decay{Z}{\ccbar}, hadronic $W$ boson decays, and top quark decays are used to investigate the properties of these resonant and quasi-resonant backgrounds. All simulated samples have been generated with \pythia~\cite{Sjostrand:2007gs,*Sjostrand:2006za} at leading order with the CT09MCS PDF set. The multijet QCD background, which is challenging to simulate accurately, is estimated using a data-driven method based on the control region defined in Sec.~\ref{sec:sr_cr}.

\subsection{Event selection}
\label{sec:EventSelection}

Events that pass the trigger criteria for heavy-flavour dijets are further selected by reconstructing two jets originating from the same PV. Kinematic requirements are applied to the reconstructed jets. Both jets are required to have $\pt>20 \gevc$ and $2.2 < \eta < 4.2$. 
The kinematic selection includes a requirement on the difference in the azimuthal angle between the jets, $|\Delta \phi|>1.5$ rad.
In 0.4$\%$ of the selected events, multiple dijet candidates exist after applying all the requirements; in such cases, the pair with the largest scalar sum of jet transverse momenta is selected. A total of 24.4 million candidates meet these criteria.

\subsection{Definition of signal and control regions}
\label{sec:sr_cr}

The main challenge in the search for $H \to \bbbar$ and $H \to \ccbar$ decays is to accurately model the multijet QCD background. Simulations are insufficient to reliably describe this background component; hence, a data-driven approach is employed.
The outputs of the tagging DNN, evaluated on the two selected jets, are used to define the SR and the CR.

To define the SR for the $H \to \bbbar$ ($H \to \ccbar$) search, specific requirements are applied to the DNN probabilities to suppress contamination from $\cquark$-jets ($\bquark$-jets) and light-flavour jets. These requirements are optimised to maximise the expected signal significance.

The CR is defined as the region in which only one of the two jets satisfies the SR DNN criteria, while a looser or orthogonal selection is applied to the other jet. This configuration predominantly selects dijet events with mixed jet flavours. Since no significant contributions from SM resonances are expected in this region, it is used to characterise the properties of the multijet QCD background. The CR selection on the second jet is designed to ensure sufficient statistical power while minimising signal and resonance contamination.

For the $H \to \bbbar$ search, the CR is derived from a dataset collected using a trigger line that requires only one secondary vertex per event, subject to a prescale, \textit{i.e.} by only recording a fixed fraction of the events. The SR requirement is applied to the jet containing the secondary vertex, while the CR requirement is applied to the second jet, regardless of whether it contains a secondary vertex. This strategy aims to reduce resonance contamination in the CR. In contrast, for the $H \to \ccbar$ search, the CR is constructed from a dataset in which both jets are associated with reconstructed secondary vertices, resulting in a higher purity sample.

The full set of requirements used to define the SR and CR is summarised in Table~\ref{tab:sr_cr_cuts}. The table also reports the \textit{purity ratio} ($R_{\rm p}$), used as a figure of merit to quantify the separation between the SR and CR. It is defined as
\begin{equation}
R_{\rm p} \equiv \frac{N^{\mathrm{SR}}_{\mathrm{sig}}/N^{\mathrm{SR}}_{\mathrm{data}}}{N^{\mathrm{CR}}_{\mathrm{sig}}/N^{\mathrm{CR}}_{\mathrm{data}}},
\end{equation}
where $N^{\mathrm{SR}}_{\mathrm{sig}}$ ($N^{\mathrm{CR}}_{\mathrm{sig}}$) denotes the expected number of signal events in the SR (CR), and $N^{\mathrm{SR}}_{\mathrm{data}}$ ($N^{\mathrm{CR}}_{\mathrm{data}}$) represents the corresponding number of data events in SR (CR).

\begin{table}[!tb]
    \centering
    \caption{Requirements applied to dijet candidates in the SR and CR samples and purity ratio. The secondary-vertex tagging requirement is referred to as ``SV-tag". $P_b$ and $P_c$ are the DNN outputs for a jet to be respectively a $b$- or $c$-jet.}
    \begin{tabular}{c|c|c}
     Search  & $H \to \bbbar$ &  $H \to \ccbar$ \\
     \hline
     SR requirements & \footnotesize{$P_{b}>0.25$ and SV-tag on both jets} & \footnotesize{$P_{c}>0.15$ and SV-tag on both jets} \\
     \hline
     CR requirements & \footnotesize{$P_{b}>0.25$ and SV-tag on one jet} & \footnotesize{$P_{c}>0.15$ and SV-tag on one jet} \\
                      & \footnotesize{$P_{b}<0.25$ on the other jet} & \footnotesize{$P_{c}<0.02$ and SV-tag on the other jet} \\
    \hline
     $R_{\rm p}$ & 5.5 & 35 \\
    \end{tabular}
    \label{tab:sr_cr_cuts}
\end{table}

\subsection{Selection efficiency and corrections}
\label{sec:corrections}

Several potential sources of inefficiencies must be accounted for in this analysis in order to obtain reliable estimations for the signal and background yields. These include the secondary-vertex tagging efficiency, the DNN selection efficiency, and the efficiency of the GEC. These efficiencies are primarily evaluated using the simulation samples described in Sec.~\ref{sec:DataSim}. However, differences between data and simulation can affect both the event selection efficiency and the shape of the dijet invariant-mass distribution. To mitigate these discrepancies, dedicated correction procedures are applied to the simulated samples.

The secondary-vertex tagging efficiency is known to differ between data and simulation, as seen in previous LHCb analyses, particularly in Ref.~\cite{LHCb-PAPER-2015-016}. To correct for this mismatch, a reweighting procedure established in 2010--2012 data, and validated in the 2015--2016 data~\cite{LHCb-PAPER-2017-050}, is applied. Events in simulation are weighted as a function of the jet \pt to reproduce the tagging efficiency observed in data.

The efficiency of the DNN selection is also subject to differences between data and simulation. A tag-and-probe method is used to derive correction factors. Dijet candidates are selected according to the preselection described in Sec.on~\ref{sec:sr_cr}. The tag jet is identified by applying a tight selection on the DNN output ($P_b$ or $P_c$), while the probe jet is selected with a requirement on the azimuthal separation from the tag jet, $|\Delta\phi| > 2.7$ rad, yielding a sample enriched in $\bquark$- or $\cquark$-jets. The efficiency of the DNN requirement on the probe jet is then measured in data as a function of \pt and compared to the corresponding efficiency in simulation. The ratio of these efficiencies is applied as a weight to the simulated events.

The GEC efficiency observed in the simulation is typically higher than the GEC efficiency in the data. To correct for this, the distribution of the number of hits in the scintillating-pad detector is measured in data using events selected with prescaled triggers that apply loose requirements. This distribution is then compared to the one obtained in simulation, and scale factors are used to reweight the simulation to correct the mismatch in event multiplicity. The corrected GEC efficiency is computed by applying the hardware-level nSPD requirement to the reweighted events.

After applying all corrections, the expected $H \to \bbbar$, $H \to \ccbar$, $Z \to \bbbar$, and $Z \to \ccbar$ yields in the signal regions are extracted, as summarised in Table~\ref{tab:expected}. Contributions from $W$ boson and top quark decays are not explicitly included in the dijet invariant-mass model, as their yields are subdominant. Their effect is instead treated as a source of systematic uncertainty.

\begin{table}[!tb]
    \centering
\caption{Expected number of SM $\bbbar$ and $\ccbar$ resonances in the SR, and number of selected data candidates.}
        \begin{tabular}{c|c|c}
     & $H \to \bbbar$ search &  $H \to \ccbar$ search \\
    \hline
     $H \to \bbbar$ & 239 & 23 \\
     $H \to \ccbar$ & 0.53 & 0.80 \\
     $Z \to \bbbar$ & 107 920 &  9 568 \\
     $Z \to \ccbar$ & 1 391 & 2 379 \\
     \hline
     data yield & 21M & 4.1M  \\

    \end{tabular}
    \label{tab:expected}
\end{table}

\subsection{Fit to the dijet invariant-mass distribution}
\label{subsec:Fit}

A binned maximum-likelihood approach is employed to fit the Higgs boson invariant mass and determine the corresponding upper limits. The observable used in the fit is the dijet invariant mass ($M_{jj}$). Invariant-mass distributions in the fit model are represented as templates, \ie histograms with 41 bins of equal width spanning the $M_{jj}$ range \mbox{[45, 250]~\gevcc}. Candidates falling outside this range are excluded from the analysis.

The fit model is defined as
\begin{align}
\label{eq:fit_model}
f(x) &= N_{\mathrm{QCD},\mathrm{SR}} \cdot B_{\mathrm{CR}}(x) \cdot \mathrm{TF}(x) + N_{Hbb,\mathrm{SR}} \cdot B_{Hbb}(x) \nonumber \\ &+ N_{Hcc,\mathrm{SR}} \cdot B_{Hcc}(x) + N_{Zbb,\mathrm{SR}}^{\rm SM} \cdot R_Z \cdot B_{Zbb}(x) + N_{Zcc,\mathrm{SR}}^{\rm SM} \cdot R_Z \cdot B_{Zcc}(x),
\end{align}
where:
\begin{itemize}
\item The variable $x$ denotes the dijet invariant mass $M_{jj}$; 
\item The functions $B_{Hbb}(x)$, $B_{Hcc}(x)$, $B_{Zbb}(x)$, and $B_{Zcc}(x)$ are the binned probability density functions (PDFs) for the $H \to \bbbar$, $H \to \ccbar$, $Z \to \bbbar$, and $Z \to \ccbar$ processes, respectively. These templates are derived from simulations after applying the SR selection; 
\item The terms $N^{\rm SM}_{Zbb,\mathrm{SR}}$ and $N^{\rm SM}_{Zcc,\mathrm{SR}}$ represent the expected $Z \to \bbbar$ and $Z \to \ccbar$ yields in the SR, which are fixed in the fit. The normalisation of the $Z$ boson contributions is scaled by the free parameter $R_Z$, which scales both $Z \to \bbbar$ and $Z \to \ccbar$ components equally;
\item The parameters $N_{Hbb,\mathrm{SR}}$ and $N_{Hcc,\mathrm{SR}}$ correspond to the number of Higgs boson candidates in the SR for the $H \to \bbbar$ and $H \to \ccbar$ decay channels, respectively. In the $H \to \bbbar$ search, the $H \to \ccbar$ yield is fixed to its SM prediction, and vice versa in the $H \to \ccbar$ search. This introduces a systematic uncertainty, which is further discussed in Sec.~\ref{sec:SystematicUncertainties}; \item The number of QCD multijet background events in the SR is denoted by $N_{\mathrm{QCD},\mathrm{SR}}$ and is treated as a free parameter in the fit;
\item The QCD background is modelled as the product $B_{\mathrm{CR}}(x) \cdot \mathrm{TF}(x)$, where $B_{\mathrm{CR}}(x)$ is the dijet mass distribution observed in the control region, and $\mathrm{TF}(x)$ is a transfer function that extrapolates the CR shape to the background in the SR. The transfer function ($\mathrm{TF}$) is defined by a Bernstein polynomial as
\begin{equation}
\label{eq:tf}
\mathrm{TF}(x) = N \cdot \sum_{i=0}^{n} c_i P_{i,n}(x) \ ,
\end{equation}
where $N$ is a normalisation factor, $c_i$ are positive coefficients, and $P_{i,n}(x)$ are the Bernstein basis polynomials of degree $n$, given by
\begin{equation}
P_{i,n}(x) = \binom{n}{i} x^i (1 - x)^{n - i}.
\end{equation}
For normalisation purposes, the coefficient $c_0$ is fixed to 1.
\end{itemize}

The fit to the dijet invariant-mass distribution is used to simultaneously determine the Higgs boson yield, the normalisation of the $Z$ boson processes, and the $\mathrm{TF}$ parameters, along with the order of the Bernstein polynomials for the $\mathrm{TF}$ optimised by selecting the configuration that yields the best goodness-of-fit, as measured by $\chi^2/\mathrm{ndof}$. The fit is performed iteratively: it begins with the Higgs yield fixed to zero and the $Z$ boson normalisation set to its SM expectation, allowing the $\mathrm{TF}$ parameters to be initialised. A second step varies the $Z$ normalisation, and finally, all components, including the Higgs boson yield, are allowed to vary in the fit. In the $H \to \ccbar$ search, the $Z$ normalisation is fixed to the value obtained from the $H \to \bbbar$ fit.

To validate the procedure and assess potential biases, 1000 pseudoexperiments are generated from the full fit model, using $\mathrm{TF}$ parameters obtained from data and SM-normalised resonance contributions. The fits are repeated on each pseudodataset to evaluate the pull distribution, defined as the difference between the fitted and expected Higgs yield, normalised to the expected yield. Small biases are observed and used to correct the final fit results.

\subsection{Determination of upper limits}
\label{subsec:UpperLimit}

Upper limits are derived using the CL$_\mathrm{s}$ method~\cite{CLs}.
The CL$_\mathrm{s}$ values are obtained by performing fits in which the number of QCD background events is allowed to vary for different assumed Higgs boson yields. Systematic uncertainties are incorporated into the fit through the inclusion of nuisance parameters, as described in Sec.~\ref{sec:SystematicUncertainties}.

The $95\%$ confidence level upper limits on the Higgs boson yields are then translated into upper limits on the cross-sections using the relation
\begin{equation}
\sigma_{\mathrm{UP}} = \sigma_{\mathrm{SM}} \cdot \frac{N_{\mathrm{UP}}}{N_{\mathrm{SM}}},
\end{equation}
where $N_{\mathrm{UP}}$ is the upper limit on the number of Higgs candidates, $N_{\mathrm{SM}}$ is the expected number of Higgs candidates under the SM, $\sigma_{\mathrm{UP}}$ is the upper limit on the cross-section, and $\sigma_{\mathrm{SM}}$ is the SM cross-section.
In addition, the expected CL$_\mathrm{s}$ values under the background-only hypothesis and the expected upper limits are evaluated. 

\subsection{Systematic uncertainties}
\label{sec:SystematicUncertainties}

A list of the sources of systematic uncertainty considered in this study is given in Table~\ref{tab:systematics_overview}. To evaluate the impact of each systematic uncertainty, the expected upper limit is first computed with all nuisance parameters included in the fit. Then, each nuisance parameter is removed individually, and the limit is recalculated. The relative change between the baseline limit and the one obtained without a specific nuisance parameter is taken as the impact of the systematic source associated with that nuisance parameter.

\begin{table}[!b]
  \begin{center}
\caption{\label{tab:systematics_overview} Systematic uncertainties considered in the determination of the expected upper limit on the $H \to \bbbar$ and  $H \to \ccbar$ cross-sections. The quoted values, in percent, represent the relative change between the baseline limit and the one obtained without the specific nuisance parameter.}      
\begin{tabular}{c|c|c}
        Systematic source &  $H \to \bbbar$ search & $H \to \ccbar$ search \\
        \hline
       GEC & {\phantom{1}}0.4$\%$ & \phantom{1}0.1$\%$  \\
        Trigger & \phantom{1}3.2$\%$  & \phantom{1}3.6$\%$ \\
        Jet secondary-vertex tagging & 15.0$\%$  & 19.0$\%$  \\
        Jet identification & \phantom{1}0.7$\%$ & \phantom{1}0.1$\%$   \\
        Jet-energy scale & \phantom{1}1.2$\%$ & \phantom{1}0.3$\%$   \\
        Jet-energy resolution & $<0.1\%$\phantom{1} & \phantom{1}0.2$\%$  \\
        Background model& \phantom{1}$1.0\%$ & \phantom{1}$1.2\%$  \\
        $Z$ boson yield & \phantom{1}2.0$\%$ & \phantom{1}1.0$\%$\\
        DNN & \phantom{1}0.1$\%$  & \phantom{1}1.8$\%$    \\
        SM backgrounds & \phantom{1}1.8$\%$ & \phantom{1}1.7$\%$ \\
        \hline
        Luminosity & \phantom{1}1.2$\%$  & \phantom{1}0.2$\%$ \\
      \end{tabular}
  \end{center}     
\end{table}

Simulation samples are corrected for differences between data and simulation of the GEC efficiencies (Sec.~\ref{sec:corrections}).
The GEC uncertainty is evaluated as the variation from the mean efficiency values, corresponding to $\pm 1$ standard deviations, and is considered as a systematic uncertainty for the normalisation of the templates.
Considering the GEC uncertainty, the upper limits increase by 0.4$\%$ for the $H \to \bbbar$ and by 0.1$\%$ for the $H \to \ccbar$ decay.

The systematic uncertainty on the trigger efficiency, except for the GEC and the secondary-vertex reconstruction, is obtained by applying a tag-and-probe technique using $Z$+jet data and simulation samples.
The tag is the reconstructed $Z$ boson in the dimuon final state, with at least one muon passing all the stages of the electroweak muon online selection, with the dedicated trigger selection where the GEC is loose. The $Z$ boson and the probe jet must be back-to-back in the azimuthal plane with $|\Delta \phi|>2.3$ rad.
The efficiency of the hardware trigger and the first software level trigger is computed on the probe jet and compared between data and simulation. The efficiencies are compatible across different intervals of jet \pt, thus no correction is needed to account for differences between data and simulation, and the statistical uncertainty on the tag-and-probe sample is taken as systematic uncertainty. The trigger uncertainty is expected to increase the $H \to \bbbar$ and $H \to \ccbar$ cross-section expected upper limits by $3.2\%$ and $3.6\%$ respectively.

The secondary-vertex tagging systematic uncertainty is related to the calibration explained in Ref.~\cite{LHCb-PAPER-2015-016} in which these values are computed with data taken at $\sqrt{s}= 7$ and 8 \tev, and applied to Run~2 data as discussed in Ref.~\cite{LHCb-PAPER-2017-050}. In those studies, a tag-and-probe technique was used on control samples with a jet and a $W$ boson, or a $B$ or $D$ meson to obtain weights for correcting the secondary-vertex tagging efficiency in simulation. These corrections and uncertainties are verified using those in Ref.~\cite{LHCb-PAPER-2017-050} to agree within 3\% between that data sample and that used in this analysis.
These weights are then applied to the reconstructed jets of the simulation samples, as a function of the jet \pt, in order to estimate the expected Higgs and $Z$ boson yields and corresponding templates. The uncertainty is propagated to the template normalisation and shapes obtained from the simulation through the corresponding weights. This systematic uncertainty is expected to increase the $H \to \bbbar$ and $H \to \ccbar$ upper limits by $15\%$ and $19\%$ respectively, and it is the dominant systematic uncertainty for both searches.

A systematic uncertainty arises from the mismodelling of jet identification variables, affecting the jet selection efficiency. The same $Z$+jet data and simulation samples described above are used to determine the uncertainty. As described in Ref.~\cite{LHCb-PAPER-2013-058}, the uncertainty is obtained by tightening these requirements and comparing the fraction of events rejected in data and simulation. The maximum relative variation is obtained by varying the maximum fraction of transverse momentum carried by a single particle inside the jet, and this is taken as the uncertainty. The expected upper limit is found to be increased by this source of systematic uncertainty by 0.7\% and $<0.1\%$ for the $H\to \bbbar$ and $H\to \ccbar$ processes, respectively.

The systematic uncertainties on the jet-energy scale and jet-energy resolution depend on how well the detector response is simulated, since this affects the reconstructed jet energy in the simulation.
The $Z$+jet sample is used to determine the systematic uncertainty related to the jet-energy scale, by comparing the $\pt(\mathrm{jet})/\pt(Z)$ distributions in data and simulation, where $\pt(Z)$ is the transverse momentum of the $Z$ boson, as done in Ref.~\cite{dijet}. A scale factor is applied to \pt(jet) in simulation and varied until the \pt(jet)/\pt(Z) distribution agrees with that in data within 1 standard deviation; the corresponding variation defines the jet-energy-scale uncertainty.
The uncertainty in the jet-energy resolution is evaluated with the low-mass and high-mass sidebands of the dijet sample that are not used in the fit ($<45$ \gevcc and $>250$ \gevcc) to avoid bias. The dijet balance $\frac{\pt(\mathrm{jet1}) - \pt(\mathrm{jet2}) }{\pt(\mathrm{jet1}) + \pt(\mathrm{jet2})}$ is compared between selected data and dijet simulation. The uncertainty interval associated with the jet-energy resolution is computed as a smearing factor applied to both jet transverse momenta to agree with data by less than 1$\sigma$.
The nuisance parameters associated with the jet-energy scale and jet-energy resolution vary between the baseline result and the factors described above. They affect the dijet invariant-mass shape of the templates as well as the normalisation, since they change the fraction of jets that pass the transverse momentum threshold. 
The expected upper limits for the $H\to \bbbar$ and $H\to \ccbar$ processes are found to increase by 1.2\% and  0.3 $\%$, respectively, due to the jet-energy scale uncertainty, and by 0.03\% and  0.2\%, respectively, due to the jet-energy resolution uncertainty.

The uncertainty in the multijet QCD background modelling arises primarily from three sources: the $\mathrm{TF}$ model, the CR contamination, and the CR sample size. The $\mathrm{TF}$ model uncertainty is assessed by varying the Bernstein polynomial degree, leading to changes in the upper limits of less than 1.0\% for $H \to \bbbar$ and less than 0.05\% for $H \to \ccbar$. Contamination in the CR occurs due to residual signal and SM resonance contributions, estimated by subtracting the expected $W$, $Z$, and $H$ boson contributions, resulting in upper limit variations of $1\%$ for $H \to \bbbar$ and $1.2\%$ for $H \to \ccbar$. The uncertainty related to the CR sample size is evaluated through bootstrap resampling, contributing approximately $0.4\%$ to both $H \to \bbbar$ and $H \to \ccbar$ limits. Overall, the systematic uncertainty from multijet QCD background modelling increases both limits by about $1\%$.

The systematic uncertainty related to the DNN selection efficiency is evaluated by considering the contamination of the probe jet samples with light and $\cquark$-quark and $\bquark$-quark flavours. Since the secondary-vertex tag is required for both the probe jet and the tag jet, the contamination is expected to be low; however, a conservative estimate is made by loosening the DNN requirement on the tag jet. Following the procedure described in Sec.~\ref{sec:corrections}, new weights are obtained, and the variation between the application of the baseline weights and those on the normalisation and template shapes is considered as a systematic uncertainty.
The expected upper limit is found to increase by 0.1\% for the $H \to \bbbar$ cross-section and 1.8\% for the $H \to \ccbar$ cross-section due to this systematic uncertainty.

The normalisation of $Z$ boson processes, $Z \to \bbbar$ and $Z \to \ccbar$, is obtained from the $H \to \bbbar$ fit and fixed in the upper limit computations. The $Z \to \bbbar$ and $Z \to \ccbar$ yields are varied as nuisance parameters to take into account the statistical uncertainty on the fit result, which is approximately 6.0\%.
By applying this nuisance parameter, the expected upper limit increases by 2.0\% for the $H \to \bbbar$ cross-section and below 1.0\% for the $H \to \ccbar$ cross-section.

Finally, a systematic uncertainty is computed to account for other SM backgrounds. 
Apart from the processes considered in the fit model, subdominant contamination in the signal region can arise from $W \to q q'$ decays and top production. 
These processes are simulated and their expected yields in the $H \to \bbbar$ and $H \to \ccbar$ signal regions are determined. The $W$ boson contribution mostly overlaps with the $Z$ boson region, while the top contribution is more similar to the combinatorial background and spreads in the full mass range. 
A systematic uncertainty is assessed by repeating the evaluation of the limits after introducing the $W$ and top backgrounds as fixed contributions to the computation. The impact on the obtained limits is negligible, therefore it is not applied in the limit computation. 
The $H \to \ccbar$ ($H \to \bbbar$) contribution in the SR of the $H \to \bbbar$ ($H \to \ccbar$) search is fixed to the SM expectation. In this case, a systematic uncertainty is calculated by considering the prediction uncertainty~\cite{th_H_BR}.
A further 4.0\% variation has been summed in quadrature to the systematic uncertainty due to the cross-section uncertainty, to take into account the acceptance correction factor described in Sec.~\ref{sec:corrections}. The expected upper limit is found to increase by 1.8\% for the $H\to \bbbar$ cross-section and 1.7\% for the $H \to \ccbar$ cross-section due to this systematic uncertainty.

The luminosity determination has a precision of 2.0\%~\cite{LHCb-PAPER-2014-047}. It is implemented in the upper limit computation as a common parameter to all simulation samples, and is found to increase the expected upper limit on the $H \to \bbbar$ decay by 1.2\% and on the $H \to \ccbar$ decay by 0.2\%.

\subsection{Results for Higgs boson searches and future prospects}
\label{sec:ResultsAndProspects}

\begin{figure}[!t]
    \centering
    \includegraphics[width=0.75\textwidth]{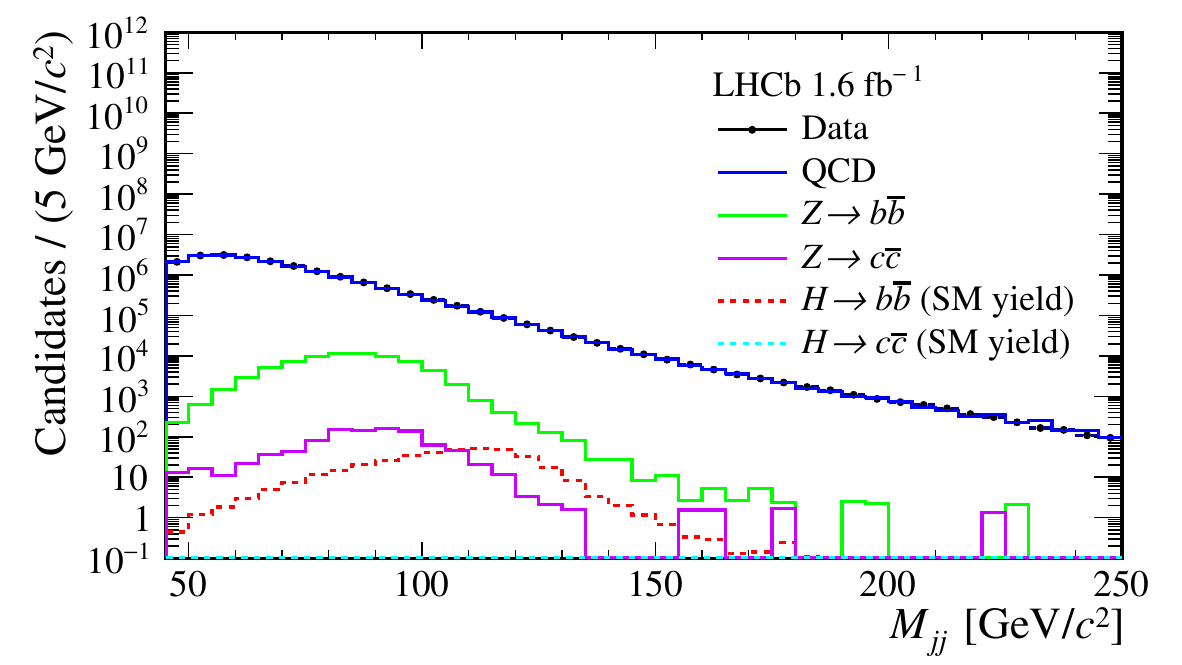}
    \includegraphics[width=0.75\textwidth]{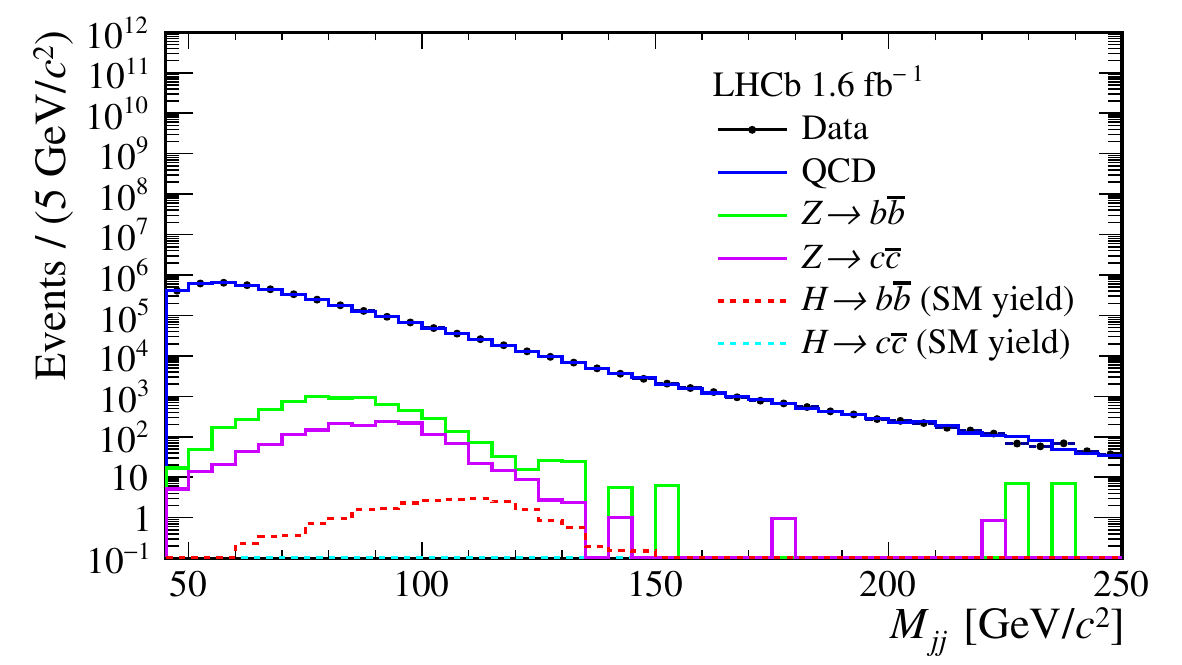}
    \caption{Dijet invariant-mass distribution with fit results for the (top) $H \to \bbbar$ search and (bottom) $H \to \ccbar$ search.}
    \label{fig:fit}
\end{figure}

The results of the fits to the dijet invariant-mass distribution in data are reported in Table~\ref{tab:fit_results}. The fitted distributions are also shown in Fig.~\ref{fig:fit}. 

In both channels, the fitted signal yields are compatible with zero within statistical uncertainties. This indicates that no significant excess attributable to a Higgs boson signal is observed in either channel. The $Z$ boson normalisation factor extracted from the fit to the $H \to \bbbar$ candidates, $R_Z = 0.655$, is consistent with the expectation from simulation within systematic uncertainties. It should be noted that an $R_Z$ value below $1$ is expected due to $Z$ boson contamination within the CR, which modifies the shape of the combinatorial dijet mass distribution. The systematic uncertainty related to this effect is considered in the limit computation, as explained in the previous section.

\begin{table}[t]
    \centering
    \caption{Results from the fit to data for the $H \to \bbbar$ and $H \to \ccbar$ searches. Quantities expressed without an uncertainty are fixed in the fit.}
    \begin{tabular}{c|c|c}
        Parameter & $H \to \bbbar$ & $H \to \ccbar$ \\
        \hline
        Signal yield & $-1900 \pm 1900$ & $-1000 \pm 1100$ \\
        $N_{\mathrm{QCD},\mathrm{SR}}$ & $(1.93 \pm 0.06) \times 10^7$ & $(3.90 \pm 0.02) \times 10^6$ \\
        $R_Z$ & $0.655 \pm 0.042$ & -- \\
        $c_0$ & 1.0 & 1.0 \\
        $c_1$ & $0.732 \pm 0.006$ & $1.390 \pm 0.006$ \\
        $c_2$ & $0.562 \pm 0.003$ & 0 \\
        $c_3$ & 0 & $2.701 \pm 0.017$ \\
        $c_4$ & $0.498 \pm 0.006$ & 0 \\
        $c_5$ & 0 & $2.9 \pm 0.2$ \\
        $c_6$ & $0.270 \pm 0.009$ & $2.2 \pm 0.3$ \\
        $c_7$ & -- & $2.3 \pm 0.4$ \\
    \end{tabular}
    \label{tab:fit_results}
\end{table}

\begin{table}[!b]
    \centering
\caption{Expected and observed upper limits on the $H \to \bbbar$ and $H \to \ccbar$ cross-sections.}
        \begin{tabular}{c|c|c|c|c|c|c}
        & \multicolumn{3}{c|}{Expected limits} & \multicolumn{3}{c}{Observed limits} \\
        \hline
       Process  & $\sigma_{\rm UP}/\sigma_{\rm SM}$& N$_{\rm UP}$ & $\sigma_{\rm UP}$ [pb] & $\sigma_{\rm UP}/\sigma_{\rm SM}$& N$_{\rm UP}$ & $\sigma_{\rm UP}$ [pb] \\
       \hline
        $H  \to \bbbar$  & 11.1 & 2652 & 353 & 6.64 & 1587 & 211 \\
        $H  \to \ccbar$  & 1834 & 1414 &  2939  & 1003 & 722 & 1605 \\
    \end{tabular}
    \label{tab:uplim}
\end{table}
\begin{figure}[!tb]
  \begin{center}
    \includegraphics[width=0.75\linewidth]{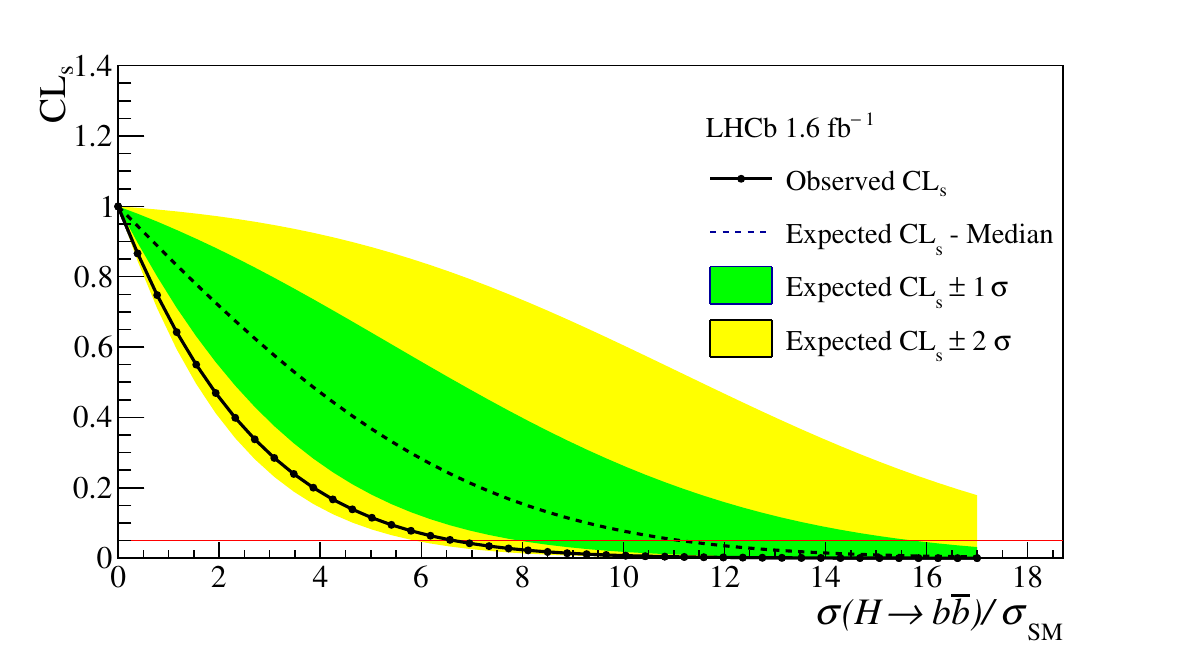}
     \includegraphics[width=0.75\linewidth]{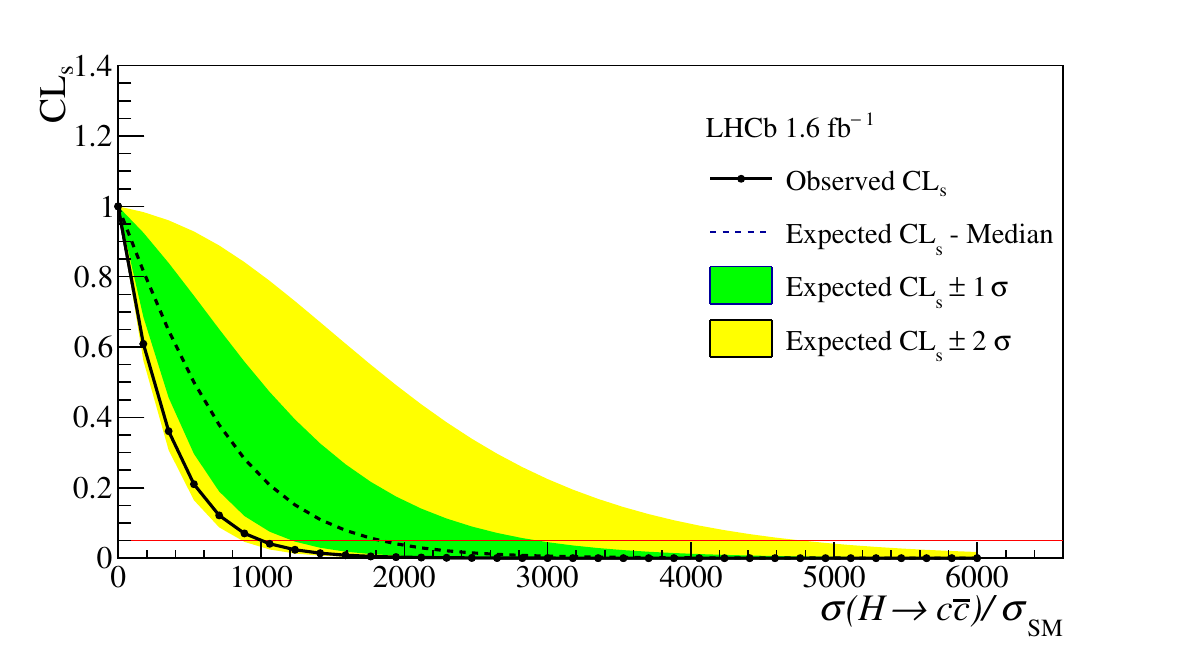}
  \end{center}
  \caption{Observed and expected $95\%$ CL upper limits on the (top) $H \to \bbbar$ and (bottom) $H \to \ccbar$ cross-sections. The red lines correspond to the $95\%$ CL upper limit.}
  \label{final_upper_limit_hcc}
\end{figure}

The corresponding 95\% CL upper limits on the production cross-section times branching fraction, including systematic uncertainties, are summarised in Table~\ref{tab:uplim}. The expected and observed limits are also shown in Fig.~\ref{final_upper_limit_hcc}.
Limits are presented in the fiducial region defined by two $\bquark$ (or $\cquark$) quarks from the Higgs decay with pseudorapidity between 2 and 5. The limits are also reported normalised to the SM cross-section.

For the $H \to \bbbar$ channel, the 95\% CL observed upper limit is $\sigma_{\mathrm{UP}} = 211$~pb, corresponding to $6.64 \times \sigma_{\mathrm{SM}}$. This is more stringent than the expected limit of $11.1 \times \sigma_{\mathrm{SM}}$, due to a downward statistical fluctuation in the data.

In the $H \to \ccbar$ channel, the observed 95\% CL upper limit is $\sigma_{\rm UP} = 1605$~pb, or $1003 \times \sigma_{\mathrm{SM}}$, while the expected limit is $1834 \times \sigma_{\mathrm{SM}}$. The sensitivity in this channel is degraded due to a low signal-to-background ratio, reduced charm-tagging efficiency, and increased background misidentification rates. These aspects reflect the experimental difficulty of probing $H \to \ccbar$ decays.

The negative yield results from the fits are associated with an underfluctuation of both $H \to \bbbar$ and $H \to \ccbar$ limits. However, it should be noted that most systematic uncertainties are correlated between the two limits, in particular, the dominant one related to secondary-vertex tagging.

Overall, no evidence of Higgs boson decays to either $\bbbar$ or $\ccbar$ quark pair is observed. The results from the $\bbbar$ search approach the SM expectation within an order of magnitude, while the limits for the $\ccbar$ final state remain substantially weaker. These findings set the most stringent limits with the current LHCb dataset and provide a benchmark for future LHCb analyses with larger datasets and improved flavour-tagging performance.

Several developments are expected to enhance the sensitivity of future searches significantly. 
An extrapolation is performed, based on assumptions driven by detector upgrades and analysis improvements, as well as increased statistics accumulated during the ongoing (Run~3) and future (Run~4 and 5) higher-luminosity run periods~\cite{LHCb-DP-2022-002, LHCb-TDR-026, LHCb-PUB-2025-001}.
Additional assumptions include improved trigger efficiency and enhanced tracking from the upgraded detector, affecting both tagging efficiency and systematic uncertainties.
The extrapolated limits on the $H \to \bbbar$ and $H \to \ccbar$ cross-sections, along with projections on the charm Yukawa coupling $y_c/y_c^{\rm SM}$, are reported in Tables~\ref{tab:extrapolation} and~\ref{tab:extrapolation_y}, respectively.

\begin{table}[!b]
    \centering
\caption{Extrapolation of the expected 95\% CL upper limits on the Higgs boson cross-section times branching ratio with datasets up to and including Run~4 and Run~5. }
        \begin{tabular}{c|c|c|c}
       & $\sigma_{\rm UP}/\sigma_{\rm SM}$  & $\sigma_{\rm UP}/\sigma_{\rm SM}$  & $\sigma_{\rm UP}/\sigma_{\rm SM}$  \\
       
       & 2016, 1.6\invfb & Runs 1-4, 50\invfb & Runs 1-5, 300\invfb \\
       \hline
       $H \to $  $ \bbbar$ &  11.1 &  1.1 & 0.38 \\
        $H \to \ccbar$ &  1834 & 141 & 45 \\
    \end{tabular}
\label{tab:extrapolation}
\end{table}
\begin{table}[!htb]
    \centering
\caption{Extrapolation of charm Yukawa coupling $y_c/y_c^{\rm SM}$ 95\% CL upper limits up to and including Run~4 and Run~5. The changes in the Higgs width due to large charm Yukawa couplings are ignored for simplicity.}
\begin{tabular}{c|c|c}
        $y_c/y_c^{\rm SM}$ &$y_c/y_c^{\rm SM}$ &  $y_c/y_c^{\rm SM}$  \\
       
        2016, 1.6\invfb & Runs 1-4, 50\invfb & Runs 1-5, 300\invfb \\
       \hline
         43 & 12 & 6.7 \\
         
    \end{tabular}
    
    \label{tab:extrapolation_y}
\end{table}

From these extrapolations, it can be concluded that the observation of the inclusive $H \to \bbbar$ process at the LHCb experiment is expected to be feasible with the dataset collected by the end of Run~4.
The upper limit on $y_c$ achievable with the inclusive $H \to \ccbar$ channel by the end of Run~5 can be compared with the limit from the $VH$ production channel reported in Ref.~\cite{yellow_paper}, which is also expected to be reached at LHCb by the end of Run~5 and is estimated to constrain $y_c$ to 2--3 times its SM value. However, the $VH$ analyses rely on assumptions about the production mechanism, while the inclusive approach described here, dominated by the gluon fusion process, is not affected by this constraint. In the SM context, the $VH$ and inclusive channels could be combined to improve overall sensitivity. 

\section{Summary}
\label{sec:Summary}
In this paper, two machine learning techniques have been developed and applied to data collected by the LHCb experiment for the first time: a regression technique designed to improve the dijet invariant-mass resolution, and a jet flavour identification algorithm based on deep neural networks. The gradient-boosted regression method exploits a comprehensive set of jet substructure variables to derive per-jet-energy corrections, significantly improving the resolution of the dijet invariant mass with respect to the standard correction procedure. The DNN jet tagging algorithm, inspired by the \emph{DeepJet} approach, combines charged particles, neutral particles, secondary-vertex, and global features to achieve a substantial gain in tagging efficiency for both $\bquark$- and $\cquark$-jets compared to the established secondary-vertex tagging algorithm at LHCb. The deployment of these techniques demonstrates the potential of modern machine learning methods to enhance the performance of jet measurements at LHCb, thereby improving the precision of analyses involving heavy-flavour jets.

As a key application of these methods, upper limits on the inclusive production of $H \to \bbbar$ and $H \to \ccbar$ decays have been set using the 2016 dataset. The observed (expected) 95\% confidence level upper limits correspond to 6.6 (11) times the SM cross-section for the $H \to \bbbar$ process, and 1003 (1834) times the SM cross-section for the $H \to \ccbar$ process.

%% file: acknowledgements.tex
\section*{Acknowledgements}
%
%
\noindent We express our gratitude to our colleagues in the CERN
accelerator departments for the excellent performance of the LHC. We
thank the technical and administrative staff at the LHCb
institutes.
We acknowledge support from CERN and from the national agencies:
ARC (Australia);
CAPES, CNPq, FAPERJ and FINEP (Brazil); 
MOST and NSFC (China); 
CNRS/IN2P3 (France); 
BMFTR, DFG and MPG (Germany);
INFN (Italy); 
NWO (Netherlands); 
MNiSW and NCN (Poland); 
MCID/IFA (Romania); 
MICIU and AEI (Spain);
SNSF and SER (Switzerland); 
NASU (Ukraine); 
STFC (United Kingdom); 
DOE NP and NSF (USA).
We acknowledge the computing resources that are provided by ARDC (Australia), 
CBPF (Brazil),
CERN, 
IHEP and LZU (China),
IN2P3 (France), 
KIT and DESY (Germany), 
INFN (Italy), 
SURF (Netherlands),
Polish WLCG (Poland),
IFIN-HH (Romania), 
PIC (Spain), CSCS (Switzerland), 
and GridPP (United Kingdom).
We are indebted to the communities behind the multiple open-source
software packages on which we depend.
Individual groups or members have received support from
Key Research Program of Frontier Sciences of CAS, CAS PIFI, CAS CCEPP, 
Minciencias (Colombia);
EPLANET, Marie Sk\l{}odowska-Curie Actions, ERC and NextGenerationEU (European Union);
A*MIDEX, ANR, IPhU and Labex P2IO, and R\'{e}gion Auvergne-Rh\^{o}ne-Alpes (France);
Alexander-von-Humboldt Foundation (Germany);
ICSC (Italy); 
Severo Ochoa and Mar\'ia de Maeztu Units of Excellence, GVA, XuntaGal, GENCAT, InTalent-Inditex and Prog.~Atracci\'on Talento CM (Spain);
SRC (Sweden);
the Leverhulme Trust, the Royal Society and UKRI (United Kingdom).

%% file: Authorship_LHCb-PAPER-2025-034.tex
\centerline
{\large\bf LHCb collaboration}
\begin
{flushleft}
\small
R.~Aaij$^{38}$\lhcborcid{0000-0003-0533-1952},
A.S.W.~Abdelmotteleb$^{57}$\lhcborcid{0000-0001-7905-0542},
C.~Abellan~Beteta$^{51}$\lhcborcid{0009-0009-0869-6798},
F.~Abudin{\'e}n$^{57}$\lhcborcid{0000-0002-6737-3528},
T.~Ackernley$^{61}$\lhcborcid{0000-0002-5951-3498},
A. A. ~Adefisoye$^{69}$\lhcborcid{0000-0003-2448-1550},
B.~Adeva$^{47}$\lhcborcid{0000-0001-9756-3712},
M.~Adinolfi$^{55}$\lhcborcid{0000-0002-1326-1264},
P.~Adlarson$^{85}$\lhcborcid{0000-0001-6280-3851},
C.~Agapopoulou$^{14}$\lhcborcid{0000-0002-2368-0147}
% ...

{\footnotesize \it

% ...
$^{14}$Universit{\'e} Paris-Saclay, CNRS/IN2P3, IJCLab, Orsay, France\\
% ...
$^{38}$Nikhef National Institute for Subatomic Physics, Amsterdam, Netherlands\\
% ...
$^{47}$Instituto Galego de F{\'\i}sica de Altas Enerx{\'\i}as (IGFAE), Universidade de Santiago de Compostela, Santiago de Compostela, Spain\\
% ...
$^{51}$Physik-Institut, Universit{\"a}t Z{\"u}rich, Z{\"u}rich, Switzerland\\
% ...
$^{55}$H.H. Wills Physics Laboratory, University of Bristol, Bristol, United Kingdom\\
% ...
$^{57}$Department of Physics, University of Warwick, Coventry, United Kingdom\\
% ...
$^{60}$School of Physics and Astronomy, University of Glasgow, Glasgow, United Kingdom\\
$^{61}$Oliver Lodge Laboratory, University of Liverpool, Liverpool, United Kingdom\\
% ...
$^{69}$Syracuse University, Syracuse, NY, United States\\
% ...
$^{85}$Department of Physics and Astronomy, Uppsala University, Uppsala, Sweden, associated to $^{60}$\\
% ...
}
\end{flushleft}